\documentclass[prd,aps,eqsecnum,floatfix,nofootinbib,preprint,tightenlines]{revtex4}
\usepackage{latexsym}
\usepackage{amsmath}
\usepackage{hyperref}

\def\NON{\nonumber\\}
\def\bibi{\bibitem}

\let\und=\b                     

\def\a{\alpha}

\def\c{\chi}
\def\d{\delta}
\def\e{\epsilon}                
\def\f{\phi}                    
\def\g{\gamma}
\def\h{\eta}

\def\j{\psi}

\def\l{\lambda}
\def\m{\mu}
\def\n{\nu}

\def\th{\theta}                  
\def\s{\sigma}                  

\def\x{\xi}
\def\z{\zeta}
\def\D{\Delta}
\def\F{\Phi}
\def\G{\Gamma}
\def\J{\Psi}
\def\L{\Lambda}
\def\O{\Omega}

\def\cb{{\cal B}}
\def\cc{{\cal C}}
\def\cd{{\cal D}}

\def\cf{{\cal F}}
\def\cg{{\cal G}}

\def\cj{{\cal J}}

\def\cl{{\cal L}}

\def\cn{{\cal N}}
\def\co{{\cal O}}

\def\cq{{\cal Q}}

\def\cs{{\cal S}}

\def\cw{{\cal W}}

\def\cbo{{\,\raise-.15ex\Sc [\,}}                       

\def\pa#1{\partial_{#1}}                        
\def\sl#1{\rlap{\hbox{$\mskip 1 mu /$}}#1}      
\def\Sl#1{\rlap{\hbox{$\mskip 3 mu /$}}#1}      

\def\svev#1{\left\langle #1\right\rangle}       

\def\ddt#1{{\buildrel {\hbox{\LARGE .\kern-2pt.}} \over {#1}}}

\def\ie{\mbox{\it i.e.}}
\def\eg{\mbox{\it e.g.}}
\def\etc{\mbox{\it etc.}}

\def\hc{{\rm h.c.\,}}

\def\half{{1\over 2}}
\def\Re{{\rm Re\,}}
\def\Im{{\rm Im\,}}

\def\ttl#1{{\it #1}}
\def\ttl#1{}

\def\db{\Delta}
\def\seef{{\it cf.}}
\def\wr{\cw_r}
\def\mythrm#1#2{\vskip 1ex \noindent {\bf #1.} \textit{#2.}}

\def\bl{\overline{\l}}

\def\bj{\overline{\j}}
\def\bJ{\overline{\J}}
\def\bh{\overline{\h}}

\def\bd{\overline{\d}}

\def\bK{\overline{K}}

\def\bth{\overline\theta}

\def\ts{\tilde{s}}

\def\tdb{\tilde{\db}}

\def\cbar{\bar{c}}

\def\ckdb{\check{\db}}
\def\ckcb{\check{\cb}}
\def\ckco{\check{\co}}
\def\ckQ{\check{Q}}
\def\ckG{\check{\G}}
\def\ckS{\check{S}}

\def\Scl{S_{cl}}

\def\Sext{S_{ext}}

\def\cnw{{\cn_w}}
\def\cql{{\cq_\ell}}
\def\cqgh{{\cq_{gh}}}
\def\und#1{{\underline{#1}}}
\def\ov#1{\overline{#1}}

\def\ov{\bar}
\def\slash#1{\sl{#1}}

\def\textit#1{{\it #1}\kern.1em }

\begin{document}
\hyphenation{fer-mio-nic per-tur-ba-tive pa-ra-me-tri-za-tion
pa-ra-me-tri-zed a-nom-al-ous}
\thispagestyle{empty}

\begin{center}
\vspace*{10mm}
{\large\bf Algebraic renormalization of supersymmetric gauge theories\\[5mm]
  with dimensionful parameters}
\\[12mm]
Maarten Golterman$^a$\ and\ Yigal Shamir$^b$
\\[8mm]
{\small\it
$^a$Department of Physics and Astronomy
\\San Francisco State University,
San Francisco, CA 94132, USA}
\\[5mm]
{\small\it $^b$Raymond and Beverly Sackler School of Physics and Astronomy\\
Tel-Aviv University, Ramat~Aviv,~69978~Israel}
\\[10mm]
{ABSTRACT}
\\[2mm]
\end{center}

\begin{quotation}
It is usually believed that there are no perturbative anomalies
in supersymmetric gauge theories beyond the well-known chiral anomaly.
In this paper we revisit this issue, because previously
given arguments are incomplete.  Specifically, we rule
out the existence of soft anomalies, \ie, quantum violations of
supersymmetric Ward identities proportional to a mass parameter
in a classically supersymmetric theory.
We do this by combining a previously proven theorem on the absence
of hard anomalies with a spurion analysis, using the methods of
Algebraic Renormalization.  We work in the on-shell component
formalism throughout.  In order to deal with the nonlinearity
of on-shell supersymmetry transformations, we take the spurions
to be dynamical, and show how they nevertheless can be decoupled.
\end{quotation}

\newpage
\section{\label{intro} Introduction}
Supersymmetry plays a prominent role in the
search for a more fundamental theory that underlies the Standard Model.
Phenomenological treatments usually incorporate soft supersymmetry breaking
terms into the lagrangian at the electro-weak scale
from the outset, on the premise that these
are the low-energy manifestation of spontaneous supersymmetry breaking.
The spontaneous breaking, in turn, is presumed to have occurred at some
higher energy scale, in a gauge theory that is supersymmetric
at both the classical and quantum levels.

In this paper we address the question of whether a classically
supersymmetric gauge theory can always be renormalized such that
both the gauge symmetry and supersymmetry are preserved by the quantum
theory.\footnote{
  In this paper we assume that supersymmetry is a global symmetry, and that
  the theory under consideration is power-counting renormalizable.
}
For the gauge symmetry alone, the answer is well-known:
Provided that the fermion content satisfies a certain algebraic condition---the
so called anomaly-cancellation condition---there is
no Adler--Bell--Jackiw (ABJ)
anomaly at the one-loop level, and moreover, by the Adler--Bardeen theorem
\cite{AB,BBBC},
the gauge symmetry is not anomalous to all orders in perturbation theory.
In this paper we will always assume that the anomaly-cancellation condition
is satisfied.  The question is, thus, whether supersymmetry can be preserved
simultaneously with the gauge symmetry at the quantum level.

No consistent regularization method preserves supersymmetry
and gauge invariance simultaneously.  The situation is similar
to that with chiral symmetry, where also no symmetry-preserving
regulator exists.  Indeed, this is for good reason:
chiral symmetry is anomalous, unless nontrivial
anomaly cancellation conditions are met.
Likewise, supersymmetry anomalies constitute a logical possibility that
must be studied in detail.

Working within the on-shell component formalism,
this subject was addressed using Algebraic Renormalization techniques
in Ref.~\cite{MPW1}, henceforth denoted MPW.  The main result of MPW is
the following theorem: A supersymmetric gauge theory whose lagrangian
contains only dimensionless parameters is free of supersymmetry anomalies.
In other words, there are no ``hard'' supersymmetry anomalies.
The practical implication is the following.  A consistent regularization method
(such as dimensional regularization) must be used to remove the infinities
(\eg\ by minimal subtraction) order by order in perturbation theory.
The resulting renormalized diagrams will in general fail
to preserve supersymmetry and/or (chiral) gauge invariance.  MPW's theorem
then assures us that finite, ``symmetry-restoring,'' counterterms can always
be found to restore gauge invariance and supersymmetry simultaneously
up to any given order.

This result leaves open the question of what happens
in supersymmetric gauge theories
with mass parameters, which appear through the superpotential in a  generic
supersymmetric gauge theory.  It turns out that this is a nontrivial issue:
it is possible to find operators that might occur as supersymmetry anomalies
in certain theories, including those that are relevant for constructing
supersymmetric versions of the Standard Model,
when the theory contains a $U(1)$ factor in the gauge group.
The operators that may constitute an anomalous divergence of the supersymmetry
current have the generic form
\begin{equation}
\label{example}
m^3 P_\pm\l\ ,
\end{equation}
in which $\l$ is the $U(1)$ gaugino field, $P_\pm = (1/2)(1\pm\g_5)$,
and $m^3$ stands for a product
of mass parameters of total dimension equal to three.
We will refer to examples such as Eq.~(\ref{example}) as ``soft''
anomalies, because they obviously require the presence of
mass parameters in the theory.
Apart from the presence of a $U(1)$ factor, for these soft anomalies to occur
the theory must not have charge conjugation symmetry.
In supersymmetric extensions
of the Standard Model this would exclude such anomalies at one loop,
but not at higher loops, where diagrams ``know'' about the presence of
all these ingredients.

The renormalization of supersymmetric theories with masses has been considered
before, with emphasis on the role of ``soft supersymmetry
breaking terms'' \cite{GG,MPW2,HKS}.
These terms are built from operators with mass dimension (strictly)
smaller than four, and break supersymmetry explicitly (though ``softly'')
at the classical level.  However, the possible existence of soft anomalies
in a classically supersymmetric theory has not been excluded.

In this paper we limit ourselves to the following question.
Can a massive supersymmetric gauge theory that is
exactly supersymmetric at the classical level generate an anomaly
at the quantum level that would vanish if all masses in the theory
are taken to zero?  In other words, do massive supersymmetric
theories exist in which soft explicit breaking of supersymmetry cannot
be avoided after quantization?

For massive theories that are exactly supersymmetric at the
classical level, the proof given in Ref.~\cite{MPW2} turns out to be incomplete.
Specifically,
an anomalous divergence of the supersymmetry current with the generic form
of Eq.~(\ref{example}) is not excluded by the spurion methods of Ref.~\cite{MPW2}.
This is true even though, techically, with the spurions of Ref.~\cite{MPW2}
all breaking terms are cohomologically exact.

In this paper, then, we prove that the most general supersymmetric gauge theory
in four dimensions, with arbitrary superpotential, has no anomalies to
all orders in perturbation theory, if the fermion representation satisfies the
usual ABJ anomaly-cancellation condition.  The central idea is to promote
each mass parameter $m$ in the theory to a full gauge-neutral,
chiral supermultiplet $(\f_s$, $\j_s)$ that couples to the
fields of the original theory through a Yukawa coupling $w$.
The original theory is recovered by sending $w\to 0$,
keeping $m=w\langle\f_s\rangle$ fixed.\footnote{
  Linear terms in the superpotential, if present, lead to parameters
  with mass dimension two, which, moreover, have different properties with
  respect to the symmetry transformations.  It turns out to be straightforward
  to deal with such parameters separately.
}
MPW's theorem can then be
applied to the theory that contains the spurious fields $\f_s$ and $\j_s$,
and this allows us to prove our
generalization of their theorem to massive supersymmetric gauge theories.  Since
MPW worked in the on-shell formulation, in which all fields in the theory are
physical, we make the same choice for our analysis.  This leads to certain
technical complications with regard to the use of spurion techniques.
Our solution is to promote the spurions to new dynamical fields,
thereby bringing the ``spurionized'' theory under the scope of MPW's theorem.
While at first sight this may appear unusual, it will turn out to be quite
natural to do so.  Using the mathematical techniques of filtration
we are then able to establish the desired results in the limit
that the dynamical-spurion fields are decoupled.

Our outline is as follows.  In Sec.~\ref{ARrev} we review the necessary elements
of the Algebraic Renormalization framework needed to understand both MPW's
theorem, and its application to massive supersymmetric gauge theories.  The
algebraic framework is based on the existence of a generalized BRST
operator, which, in our case, covers gauge and translation invariance,
$R$-symmetry and supersymmetry.
Any anomaly must satisfy certain algebraic conditions---the
Wess--Zumino consistency conditions \cite{WZ}.  As explained in Sec.~\ref{ARrev},
these can be formulated in the algebraic framework with the help of the
BRST operator.
In Sec.~\ref{onshell} we review the on-shell formulation of
supersymmetric gauge theories with arbitrary superpotential,
including the gauge-fixing
terms and the on-shell form of BRST transformations.  We also review MPW's
theorem, in Sec.~\ref{mpw1rev}.

Our new results are contained in Sec.~\ref{spur}, which is the main part of
this paper.
In Sec.~\ref{photino} we explain in detail why, in a massive supersymmetric
gauge theory, operators such as Eq.~(\ref{example}) are indeed candidate anomalies
that satisfy the relevant Wess--Zumino consistency conditions.
We then give an intuitive
description of our setup and proof in Secs.~\ref{outline} and \ref{dyntheory}.
Section~\ref{main} states our main result, and the rest of Sec.~\ref{spur} is
devoted to the technical details of the proof,
with Sec.~\ref{smr} giving a technical summary.
Our conclusion is contained in Sec.~\ref{conc}.

A relatively quick overview of the main ideas of our analysis can be obtained
by reading only Secs.~\ref{onshell}, the first three subsections
of Sec.~\ref{spur},
and the conclusion, while skimming elements of Sec.~\ref{ARrev}, depending on
the background of the reader.

A number of appendices take care of some technical details, as well as
some issues peripheral to the main point of this paper.
Appendix~\ref{notation} summarizes notation and conventions,
while App.~\ref{offshell} discusses the relation
of the on-shell formalism we employ in the paper to the off-shell component
formulation of the theory.  Appendix~\ref{Omega} takes care of the special
case of linear terms in the superpotential, which leads to the presence of
parameters with mass dimension two in the theory.  Appendices~\ref{embed}
and \ref{wnct} provide proofs for technical lemmas used in the proof of
Sec.~\ref{spur}.  Appendix~\ref{sspace} discusses the superspace
origin of candidate anomalies such as Eq.~(\ref{example}), showing that in
superspace they would take the shape of supergauge anomalies.

The last two appendices explain why an anomalous divergence such as
Eq.~(\ref{example}) is not ruled out by the method of Ref.~\cite{MPW2}.
In App.~\ref{ARgaugefix} we derive the continuity equation for the renormalized
supersymmetry current when the ST identity is satisfied
at the quantum level.
The derivation applies in the absence of external (spurion) fields.
In App.~\ref{MPW2spr} we contrast the appearance of anomalies in the
Algebraic-Renormalization approach with their original role of
an anomalous divergence in the continuity equation.
We conclude that the spurions introduced in Ref.~\cite{MPW2}
do allow an anomalous divergence such as Eq.~(\ref{example}),
despite the fact that, technically speaking, that anomalous divergence
becomes cohomologically trivial in the presence of these spurions.\footnote{
  A similar statement applies to the spurions introduced in Ref.~\cite{HKS}.
}

\section{\label{ARrev} Algebraic Renormalization review}
To keep the paper self-contained, this section provides a brief review
of Algebraic Renormalization.
We begin with the classical theory.
At a formal level, a euclidean quantum field theory
is defined by the partition function
\begin{equation}
  Z = \int \prod_I \cd\F_I \; \exp(-\Scl) \,.
\label{Z}
\end{equation}
Here $\F_I(x)$ stands for all fields we will be integrating over.
These include bosons, fermions, ghosts, and, possibly, auxiliary fields.
With a slight abuse of language, we will refer to them collectively
as the dynamical fields.  The classical action consists of two parts
\begin{equation}
  \Scl = S_0(\F_I(x);\z_i) + \Sext(\F_I(x),K_I(x);\z_i,k_i)\,.
\label{scl}
\end{equation}
Here $S_0$ is the usual classical action of
the theory.  The second part, $\Sext$, depends on both the dynamical fields
and on a set of external sources $K_I(x)$,
one for each dynamical field $\F_I(x)$.
For on-shell supersymmetry one has $\Sext=\Sext^{lin}+\Sext^{bil}$, where
$\Sext^{lin}$ ($\Sext^{bil}$) is linear (bilinear) in the $K$ sources.
The explicit form of the linear part is
\begin{equation}
  \Sext^{lin} = \sum_I \int d^4x\, K_I(x) s\F_I(x) + \sum_j k_j s\z_j \,.
\label{sext}
\end{equation}
The coefficient of $K_I(x)$, namely $s\F_I(x)$, is the BRST variation
of $\F_I(x)$.  Following Ref.~\cite{MPW1},
the BRST operator $s$ simultaneously encodes gauge transformations,
translations, supersymmetry and $R$-symmetry transformations
(see Sec.~\ref{onshell}).
While at this point we are still dealing with the classical theory,
the ultimate goal of introducing the source terms is
to handle nonlinear field transformations in the quantum theory.
BRST transformations make use of opposite-statistics parameters, and thus
the statistics of $s\F_I(x)$ is opposite to that of $\F_i(x)$, and the same
is true for $K_I(x)$.
In this paper, the sum on the right-hand side of Eq.~(\ref{sext})
extends over all the dynamical fields, including those that transform
linearly.\footnote{
  It will be convenient to do so, even if this differs from the
  choice made in Ref.~\cite{MPW1}.
}

In addition to fields, the classical action depends on a set of global
parameters $\z_j$.  These come in two types.
There are the lagrangian parameters that occur in $S_0$:
the gauge coupling, and the parameters of the superpotential.
In addition, there are global opposite-statistics BRST parameters
associated with translations, supersymmetry transformations, and $R$-symmetry
transformations.  Some global parameters have nontrivial
BRST transformation rules, and
we find it convenient to add to $\Sext^{lin}$ a term $k_j s\z_j$ for each global
parameter.

The transformations encoded in the BRST operator correspond to
the symmetries of the original action $S_0$, which, in turn,
is BRST invariant by construction: $sS_0=0$.\footnote{
  As described in Sec.~\ref{onshell}, dimensionful parameters that occur
  in $S_0$ will be taken to
  transform nontrivially under the $R$ symmetry, and they are thus to be
  varied as well when computing $sS_0$.
}
The trick of using opposite-statistics parameters is motivated by the goal
of having a nilpotent BRST operator, \ie,
$s^2$ vanishes when applied to any dynamical field
or parameter.  If we now extend the BRST transformation trivially to the
sources: $sK_I(x)=sk_j=0$, and set $\Sext^{bil}=0$,
it follows that the complete classical action, including the source terms,
is BRST invariant: $s\Scl=0$.
This is, in fact, the whole story for the off-shell
component formalism of supersymmetric theories.  As we will discuss below,
the situation in the on-shell formalism is more involved, and this is related to
the need to introduce bilinear  terms in the $K_I$ in $S_{ext}$.

The effective-action functional $\G=\G(\f_I(x),K_I(x);\z_i,k_i)$
is the generator of 1PI (one-particle irreducible) functions.
It depends on a
set of fields and parameters similar to those appearing in the action,
except that each dynamical field $\F_I(x)$ is traded for a corresponding
effective field $\f_I(x)$.  Of prime interest is the renormalized 1PI functional
$\G_r$, whose order by order construction is discussed below.
Its tree approximation coincides with the classical action, and we will write
\begin{equation}
  \G_r = \Scl + \G_q \,,
\label{Gq}
\end{equation}
where $\G_q$ includes all the loop corrections.

The Slavnov-Taylor (ST) operator
is a nonlinear operator acting on effective action functionals, given by
\begin{equation}
  \cs(\G) =  \sum_I \int d^4x\,
  \frac{\d\G}{\d\f_I(x)} \frac{\d\G}{\d K_I(x)}
  +\sum_j \frac{\partial\G}{\partial\z_j} \frac{\partial\G}{\partial k_j}\,.
\label{STop}
\end{equation}
For both the off-shell and on-shell supersymmetry formulations,
the classical action satisfies the ST identity, $\cs(\Scl)=0$.\footnote{
  When the ST operator acts on the classical action we make the
  natural identification $\F_I \leftrightarrow \f_I$.
}
Let us elaborate on this statement.
In the off-shell formalism it is easy to see that $\cs(\Scl)=s\Scl=0$.
The last equality has been discussed above, and requires the
nilpotency of the BRST operator $s$.
In the on-shell formalism the situation is more involved.
Once the auxiliary fields are integrated out,
$s^2$ does not vanish when applied to a fermion field.
Instead, the result is proportional to the fermion's equations of motion.
As explained in detail in App.~\ref{offshell}, the classical
on-shell action $\Scl = S_0 + \Sext^{lin} + \Sext^{bil}$
nevertheless satisfies the ST identity.

We will also need the \textit{linearized} ST operator.
For any functional $\G$, the associated linearized ST operator
\begin{equation}
  \cs_\G =  \sum_I \int d^4x\, \left(
  \frac{\d\G}{\d\f_I(x)} \frac{\d}{\d K_I(x)}
  + \frac{\d\G}{\d K_I(x)} \frac{\d}{\d\f_I(x)}
  \right)
  +\sum_j \left(\frac{\partial\G}{\partial\z_j} \frac{\partial}{\partial k_j}
  +\frac{\partial\G}{\partial k_j} \frac{\partial}{\partial \z_j} \right)\,,
\label{STlin}
\end{equation}
is an anti-commuting first-order differential operator.
Its basic properties are\footnote{
  It is straightforward to check that $\cs_\G^2$ contains no
  second-derivative terms, hence $\cs_\G^2$ is a commuting
  first-order differential operator.
  Both Eqs.~(\ref{prop1}) and~(\ref{prop2}) can be derived by working out
  the explicit expression for $\cs_\G^2$.  The relation
  $\cs(\G) = \half \cs_\G\, \G$ is also
  used in the derivation of Eq.~(\ref{prop1}).
}
\begin{equation}
  \cs_\G\, \cs(\G) = 0 \,,  \qquad \forall \G \,,
\label{prop1}
\end{equation}
and
\begin{equation}
  \cs_\G^2=0 \,,  \qquad {\rm if} \qquad \cs(\G) = 0 \,.
\label{prop2}
\end{equation}
A special role is played by the linearized
ST operator associated with the classical action,
\begin{equation}
  \cb = \cs_{\Scl} \,.
\label{cb}
\end{equation}
Since $\cs(\Scl)=0$ both on- and off-shell, it follows from
Eq.~(\ref{prop2}) that in both cases $\cb$ is nilpotent: $\cb^2=0$.
One could say that the loss of nilpotency of $s$ in the on-shell formalism is
``remedied'' by using instead the linearized ST operator $\cb$.
As we will see below, this is the nilpotent operator in terms of which
the Wess--Zumino consistency conditions are formulated.

From Eq.~(\ref{STlin}) it follows that the transformation rules of
individual fields are
\begin{equation}
  \cb \f_I(x) = \frac{\d \Scl}{\d K_I(x)}\,, \qquad\qquad
  \cb K_I(x) = \frac{\d \Scl}{\d \f_I(x)}\,.
\label{Btrans}
\end{equation}
If $\Sext^{bil}$ is independent of a particular $K_I(x)$ field,
one has $\cb \f_I(x) = s\f_I(x)$.  The same is true for all parameters:
$\cb \z_i = s\z_i$.  The exceptions are the fermion transformation rules
in the on-shell formalism.  For the explicit form of the difference
$\cb-s$ when acting on a fermion field, see Eq.~(\ref{Bs}).
Regarding the transformation rules of the source fields,
it is worth noting that $\cb K_I(x)$ always contains an inhomogeneous term
which is the (classical) equation of motion of $\f_I(x)$.

Let us turn to the quantum theory.
While the methods of Algebraic Renormalization
are largely independent of the particulars of the renormalization procedure,
to make it more concrete we will consider dimensional regularization
\cite{HV,BM}.
The first step is to extend the four-dimensional classical action
to a suitable tree-level, $d$-dimensional action
$S_{t,d}^{(0)} = S_d^{(0)} + S_{ext,d}^{(0)}$.
The subscript $t$ stands for total, \ie, it accounts for both
source-independent and source-dependent terms.
The quantum action is constructed recursively as
\begin{equation}
  S_{t,d}^{(n)} = S_{t,d}^{(n-1)} + S_{t,s}^{[n]} + S_{t,f}^{[n]} \,,
\label{ctsn}
\end{equation}
where $S_{t,d}^{(n)}$ is the action with all counterterms up to, and
including, order $n$.  There are two types of counterterms.
The $n$-th order singular counterterms, $S_{t,s}^{[n]}$,
are chosen so as to make all $n$-loop diagrams finite.  For simplicity
we will assume that $S_{t,s}^{[n]}$ corresponds to minimal subtraction.
The role of the $n$-th order finite counterterms, $S_{t,f}^{[n]}$,
is explained as follows.
Let $\G_d^{(n)}$ be the $n$-loop 1PI functional of the $d$-dimensional
theory.  After minimal subtraction at order $n$
(\ie, after adding $S_{t,s}^{[n]}$, but not yet $S_{t,f}^{[n]}$,
to $S_{t,d}^{(n-1)}$) we may take the limit $d\to 4$,
obtaining a renormalized $n$-loop 1PI functional,
\begin{equation}
  \G_r^{(n)} =  \lim_{d\to 4} \G_d^{(n)} \,.
\label{Gren}
\end{equation}

Let us introduce the \textit{breaking term}
by applying the ST operator to the $n$-th order renormalized 1PI functional
\begin{equation}
  \db^{(n)} = \cs(\G_r^{(n)}) \,.
\label{Dbreaking}
\end{equation}
Were it not for the need to regularize and renormalize a field theory,
the classical invariance could be used to infer that $\db^{(n)}$ vanishes
to all order.  In reality, some of the classical symmetries will not
be respected by the regularization.  As a result,
the minimally-subtracted 1PI functional may fail to preserve some
of the classical symmetries, and this failure is quantified by
the nonvanishing of the breaking term $\db^{(n)}$.

Assuming we have achieved $\db^{(k)}=0$ for $1\le k\le n-1$,
and that $\db^{(n)}\ne 0$ after minimal subtraction,
we use the freedom to adjust the $n$-th order finite counterterms
$S_{t,f}^{[n]}$, seeking to cancel the breaking term at order $n$ as well.
If we succeed in enforcing $\db^{(n)}=0$,
the theory is free of anomalies to this order,
and we may proceed to the next order.

Let us discuss the basic properties of the breaking term.
The \textit{Regularized Action Principle} implies that,
in the $d$-dimensional theory\footnote{
  For a proof and references to the original literature
  see \eg\ the appendix of Ref.~\cite{MSR}.
  The proof given therein simplifies considerably for
  our specific choice of the quantum transformation~(\ref{ts}).
}
\begin{equation}
  \cs(\G_d^{(n)}) = \ts_d^{(n)} S_{t,d}^{(n)} \cdot \G_d^{(n)}\,.
\label{RAP}
\end{equation}
The dot notation on the right-hand side stands for an
insertion of the variation of the quantum action, $\ts_d^{(n)} S_{t,d}^{(n)}$,
into 1PI diagrams.  The quantum transformation of the dynamical
fields is defined by
\begin{equation}
  \ts_d^{(n)} \F_I(x) \equiv \frac{\d S_{ext,d}^{(n)}}{\d K_I(x)} \,.
\label{ts}
\end{equation}
Note that, if $S_{ext}^{bil}\ne 0$, then $\ts_d^{(n)} \F_I(x)$
will contain terms that depend on $K$-source fields.
In the classical theory, \ie, if we set $n=0$ and $d=4$,
then $\ts_d^{(n)}$ reduces to the linearized ST operator $\cb$ of Eq.~(\ref{cb}),
when the latter acts on a dynamical field.

If the BRST variation of a field $s\F_I(x)$ happens to be linear in the
dynamical fields, then the operator $K_I(x) s\F_I(x)$ cannot occur as an
insertion in any 1PI diagram.  Therefore no singular counterterm
$\propto K_I(x) s\F_I(x)$ is needed, and we elect to avoid any
finite counterterms that depend on $K_I(x)$ as well.
The result is that, when $s\F_I(x)$ is linear in the dynamical fields,
\begin{equation}
  \frac{\d \G_r^{(n)}}{\d K_I(x)} = \frac{\d \Scl}{\d K_I(x)} \,,
\label{dKlin0}
\end{equation}
to all orders.  The BRST transformation of the global parameters
is similarly unrenormalized:
$\partial \G_r/\partial k_j = \partial \Scl/\partial k_j = s\z_j$.

Using the Regularized Action Principle, Eq.~(\ref{RAP}),
and the defining equation~(\ref{Dbreaking}),
we obtain the following expression for the breaking term
\begin{equation}
  \db^{(n)} = \lim_{d\to 4}\, \ts_d^{(n)} S_{t,d}^{(n)} \cdot \G_d^{(n)} \,.
\label{limDbreaking}
\end{equation}
To verify that the operator insertion on the right-hand side
is finite we may again use the Regularized Action Principle,
but now at the level
of connected functions.  For any (renormalized) operator $\co$, one has
\begin{equation}
  \svev{\ts_d^{(n)} \co}
  = \svev{\co\ \ts_d^{(n)} S_{t,d}^{(n)}} \,.
\label{RAPconn}
\end{equation}
The left-hand side is finite because it involves the renormalized
transformation $\ts_d^{(n)}$.  Therefore, the right-hand side is finite, too.

The second crucial property of the breaking term is locality.  Assume as before
that we have adjusted the symmetry-restoring (finite) counterterms
to achieve $\db^{(k)}=0$ for $k=1,2,\ldots,n-1$.
Moving on to order $n$, after performing minimal subtraction
we will in general obtain a nonzero breaking term that can
be expressed as
\begin{equation}
  \db^{(n)} = \int d^4x\, \db^{(n)}(x)\,,
\label{Dn}
\end{equation}
where $\db^{(n)}(x)$ is a \textit{local} operator.
Thanks to the vanishing of $\db^{(k)}$ for $1\le k\le n-1$,
the operator $\ts_d^{(n-1)} S_{t,d}^{(n-1)}$ is evanescent.
It follows that, at the next order,
the sum of all minimally subtracted $n$-loop
diagrams with an insertion of that evanescent operator
collapses to a contact term for $d\to 4$
(see \eg\ Ref.~\cite{JCC} and references therein).\footnote{
  Explicitly, the operator that is being inserted on the right-hand side
  of Eq.~(\ref{RAP}) at this point is
  $( \ts_d^{(n-1)} + \ts_s^{[n]} ) ( S_{t,d}^{(n-1)} + S_{t,s}^{[n]} ),$
  because no finite counterterms have been introduced at order $n$ yet.
}

The algebraic \textit{consistency conditions}
are derived by applying Eq.~(\ref{prop1}) to the minimally-subtracted
$\G_r^{(n)}$.  Ignoring terms of order $\hbar^{n+1}$ we find
\begin{equation}
  0 = \cs_{\G_r^{(n)}}\, \cs(\G_r^{(n)})
  =
  \cb  \int d^4x\, \db^{(n)}(x) \,.
\label{closed}
\end{equation}
The assumed vanishing of $\cs(\G_r^{(k)})$ for $1\le k\le n-1$
implies that $\cs(\G_r^{(n)})$ is of order $\hbar^n$, and
this allows us to replace $\cs_{\G_r^{(n)}}$ with $\cs_{\Scl}=\cb$
to the given accuracy.
In mathematical parlance, the consistency conditions~(\ref{closed})
state that $\db^{(n)}=\cs(\G_r^{(n)})$ is cohomologically closed;
it belongs to the cohomology space of the nilpotent operator $\cb$.\footnote{
  The mathematical framework is discussed in more detail in Sec.~\ref{filt}.
  Additional algebraic constraints on the breaking term
  are discussed in Sec.~\ref{onshell}.
}

The breaking term may turn out to be cohomologically exact.
This means that there exists a local operator $\cq^{(n)}(x)$ such that
\begin{equation}
  \db^{(n)}(x) = \cb \cq^{(n)}(x) + \mbox{total derivative} \,.
\label{Bexact}
\end{equation}
If we now choose the $n$-th order symmetry-restoring counterterm action as
$S_{t,f}^{[n]}=-\int d^dx\,\cq^{(n)}(x)$, the result will be a vanishing
breaking term, $\db^{(n)}=0$.  Equivalently, the quantum theory now satisfies
all the ST identities encoded in $\cs(\G_r^{(n)})=0$, up to terms of order
$\hbar^{n+1}$. At this point we have succeeded in renormalizing the theory
while preserving all its classical symmetries up to, and including, order $n$.

If there does not exist any local operator $\cq^{(n)}(x)$ that satisfies
Eq.~(\ref{Bexact}) then we have an anomaly;
starting at order $n$ it is impossible
to simultaneously satisfy all the classical symmetries in the quantum theory.

Imposing
(when possible) the vanishing of the breaking term does not uniquely determine
the finite counterterms, and the remaining freedom is fixed by a set
of renormalization conditions.

In the next section we turn to the concrete Algebraic-Renormalization
framework for on-shell supersymmetry.

\section{\label{onshell} On-shell formalism}
In this section we review the on-shell component formulation
of supersymmetric gauge theories.
Sec.~\ref{onsh} introduces the on-shell framework as adapted
to the Algebraic Renormalization methodology by MPW \cite{MPW1}.
In Sec.~\ref{mpw1rev} we review MPW's main result---the absence
of hard supersymmetry anomalies---which is
the starting point for the present work.
Appendix~\ref{notation} contains our notation, while
App.~\ref{offshell} provides a brief review of the off-shell formalism,
and elaborates on the transition
from the off-shell to the on-shell formalism.

\subsection{\label{onsh} Repository of on-shell supersymmetry}
The source-independent part of the classical action $S_0$ (see Eq.~(\ref{scl}))
is the sum of the gauge-invariant and gauge-fixing actions,
\begin{equation}
  S_0 = \int d^4x\, (\cl(x) + \cl_{egf}(x)) \,.
\label{SL}
\end{equation}
The ``extended'' gauge-fixing lagrangian $\cl_{egf}$
will be discussed later on.
The physical-field content of a supersymmetric theory consists of a set of
(on-shell) chiral multiplets $(\f_i,\j_i)$, as well as of a set
of gauge multiplets $(A_{\m a},\l_a)$, where $A_{\m a}$ denotes the gauge fields
and $\l_a$ the gauginos.  To avoid cumbersome notation we will mostly
consider a single gauge group and thus a single gauge coupling $g$.
The generalization to an arbitrary gauge group,
containing both abelian and nonabelian factors, is usually trivial.
More details on the gauge group will be discussed as needed.

The gauge-invariant, on-shell supersymmetric lagrangian is
\begin{eqnarray}
  \cl &=& {1\over 4} F_{\m\n a}^2 + {1\over 2} \bl_a \Sl{D}_{ab} \l_b
  + (D_\m \f)^*_j (D_\m \f)_j + {1\over 2} \bj_i \Sl{D}_{5ij} \j_j
  - ig\sqrt{2}\, \bl_a \f^*_{5i} T_{5aij} \j_j
\NON
  && + {g^2\over 2} \Big(\f^*_i T_{aij} \f_j \Big)^2
  + W^*_{,i}\, W_{,i} + {1\over 2} \bj_i W_{5,i,j}\, \j_j \,.
\label{onL}
\end{eqnarray}
Our notation puts together each two-component Weyl fermion and its
anti-fermion into a four-component Majorana-like spinor.
Its definition, along with the all the definitions of related objects
(such as \eg\ $\Sl{D}_5$), is given in App.~\ref{notation}.
Repeated and ``squared'' indices are summed over.
The gauge invariant superpotential $W$ has the general form
\begin{equation}
  W = \O_i \f_i + {1\over 2}M_{ij} \f_i \f_j
  + {1\over 6}Y_{ijk} \f_i \f_j \f_k \,.
\label{W}
\end{equation}
The mass dimensions of $\O_i$ and $M_{ij}$ are two and one respectively.
These dimensionful parameters play center stage in this paper.
The $Y_{ijk}$ are (dimensionless) Yukawa couplings.
$M_{ij}$ and $Y_{ijk}$ are symmetric in all indices.  The shorthand
$W_{,i} = \partial W / \partial \f_i$ \etc\ is used in Eq.~(\ref{onL}).

The BRST transformations depend
on several opposite-statistics global parameters:
anti-commuting parameters $\x_\m$ for translations and $\c$ for $R$-symmetry
transformations, and a commuting spinor $\h$ for supersymmetry transformations.
In each of the transformation rules given below,
we may split the BRST operator $s$ as
\begin{equation}
  s = s_g + s_\x + s_\h + s_\c\,.
\label{ssplit}
\end{equation}
First, $s_\c$ generates the $R$ transformations,
and contains all the $\c$-dependent terms of each transformation rule.
The $R$-symmetry is the ``canonical'' one, where all the scalar fields $\f_i$
have a common $R$-charge equal to $2/3$.\footnote{
  Any other $R$ symmetry is the sum of the canonical $R$ symmetry
  and a flavor rotation.
}
Of the remaining terms, $s_\h$ consists of the $\h$-dependent terms,
and generates the supersymmetry transformations.  The terms that are independent
of both $\c$ and $\h$ but depend on $\x_\m$ constitute the translation
part $s_\x$.  The remaining terms, that do not depend on any of the
global BRST parameters, constitute $s_g$.

We next give the explicit form of the BRST transformation rules.
The physical fields transform according to\footnote{
  Our convention is to write the part of $S_{ext}^{lin}$ that involves
  $s\j_i$ as $\bK^\j_i s\j_i$, and $\bK^\l_a s\l_a$ for
  the gauginos.
}
\begin{subequations}
\label{son}
\begin{eqnarray}
  sA_{\m a} &=& \bh\g_\m\l_a + D_{\m ab}c_b + \x_\n\pa{\n}A_{\m a} \,,
\label{sona}\\
  s\l_a &=& \Big( (i/2) F_{\m\n a}\s_{\m\n}
  -ig\g_5 \f^*_i T_{aij}\f_j \Big)\h
  + g f_{abc} c_b \l_c + \x_\n\pa{\n}\l_a + i\c \g_5 \l_a \,, \hspace{10ex}
\label{sonb}\\
  s\f_i &=& \sqrt{2}\,\bh P_+\j_i - ig c_a T_{aij}\f_j + \x_\n\pa{\n}\f_i
  +i (2/3) \c \f_i \,,
\label{sonc}\\
  s\f_i^* &=& \sqrt{2}\,\bh P_-\j_i + ig c_a T_{aij}^*\f_j^*
  + \x_\n\pa{\n}\f_i^* - i (2/3) \c \f_i^* \,,
\label{sond}\\
  s\j_i & = & \sqrt{2}\left(\Sl{D}_{5ij}^*\f_{5j}^* - W_{5,i}^*\right)\h
  - ig c_a T_{5aij}\j_j + \x_\n\pa{\n}\j_i -i(1/3)\c \g_5\j_i \,.
\label{sone}
\end{eqnarray}
\end{subequations}
The ghost-sector fields transform as
\begin{subequations}
\label{gfBRST}
\begin{eqnarray}
  sc_a &=& \bh\g_\m\h\, A_{\m a} + \frac{g}{2} f_{abc} c_b c_c
           + \x_\n\pa{\n}c_a \ ,
\label{gfBRSTa}\\
  s\cbar_a &=& -ib_a + \x_\n\pa{\n}\cbar_a \ ,
\label{gfBRSTb}\\
  sb_a &=& i\bh\g_\n\h\,\pa{\n}\cbar_a + \x_\n\pa{\n}b_a \,.
\label{gfBRSTc}
\end{eqnarray}
\end{subequations}
The parameters transform according to
\begin{subequations}
\label{paron}
\begin{eqnarray}
  s\x_\m &=& -\bh\g_\m\h \,,
\label{parona}\\
  s\h &=& i\c\g_5\h \,,
\label{paronb}\\
  s\c &=& 0 \,,
\label{paronc}\\
  s \O_i &=& i(4/3)\c \O_i \,,
\label{parond}\\
  s M_{ij} &=& i(2/3)\c M_{ij} \,.
\label{parone}
\end{eqnarray}
\end{subequations}
In Eq.~(\ref{paron}) the first three lines describe the transformation
properties of the BRST parameters themselves.  Of the lagrangian parameters,
the dimensionless ones are BRST invariant.
The last two lines of Eq.~(\ref{paron})
give the BRST transformation rules of the dimensionful superpotential
parameters.  These transformation rules promote the dimensionful
parameters to global spurions by assigning to them a nonzero
$R$-charge: $M_{ij}$ has the same charge as a scalar
field, and $\O_i$ as the product of two scalar fields.

The on-shell BRST transformations are nilpotent when applied to bosons,
ghost-sector fields, and parameters, but not when applied to fermions.
Further explanations on the structure of the on-shell classical action
and BRST transformations
may be found in App.~\ref{offshell}.  In particular we explain there how,
in spite of the failure of $s^2$ to vanish when acting on fermions,
the on-shell action nevertheless satisfies the ST identity.
We also give the explicit form of the terms that are bilinear
in the $K$-source fields, which are crucial for the validity of
the on-shell ST identity, as well as the $\cb$ transformation rules
for fermion fields, with $\cb$ being nilpotent on all fields (and sources).

Next, we turn to the gauge fixing action, given by
\begin{eqnarray}
  S_{egf}
  &=& s \int d^4x\,
  \left[ \cbar_a \left(\frac{i\a}{2}\, b_a +\cg_a \right) \right]
\NON
  &=& \rule{0ex}{4ex}   \int d^4x\,
  \left( \frac{\a}{2}\, ( b_a^2 + (\bh \g_\m \h) \cbar_a\, \pa\m \cbar_a)
  -ib_a \cg_a - \cbar_a\,(s_g+s_\h)\cg_a \right) \,.
\label{Segf}
\end{eqnarray}
Here $\a$ is the gauge parameter, and $\cg_a$ is the gauge condition,
which we will take to be the Lorenz gauge,
\begin{equation}
  \cg_a = \pa\m A_{\m a} \,.
\label{dA}
\end{equation}
Compared to its textbook form, Eq.~(\ref{Segf}) is an ``extended''
gauge-fixing action that contains extra terms coming from the application
of $s_\h$ to $b_a$ and to $\cg_a$.
If the supersymmetry--BRST parameter $\h$ is set to zero,
$S_{ext}$ reduces to the standard gauge-fixing action.
The idea behind this extended form is to maintain BRST-exactness.
Since $s$ is nilpotent when acting
on boson and on ghost-sector fields, the extended gauge-fixing action
is BRST invariant: $s S_{egf}=0$.

We will take the canonical dimension of the
ghost-sector fields $c(x)$ and $\cbar(x)$
to be one.\footnote{
  This convention is different from Ref.~\cite{MPW1}
  where the canonical dimensions of these fields are 0 and 2, respectively.
}
This implies that the mass dimension of the BRST operator is one,
and that the local breaking term $\db(x)$ has mass dimension
equal to 5.  The global BRST parameters $\x_\m$, $\h$, and $\c$,
have mass dimensions 0, $\half$, and 1, respectively.
The mass dimension of the $K$-source fields is determined from
the requirement that $S_{ext}$ be dimensionless.

Finally, we introduce the ghost number.  For the ghost field $c(x)$
as well as the global BRST parameters $\x_\m$, $\h$ and $\c$
it is equal to one. $\cbar(x)$ has ghost-number $-1$, while
the remaining dynamical fields have zero ghost number.
The ghost number of the $K$ sources is determined by the requirement
that the total action have zero ghost number.

\subsection{\label{mpw1rev} Review of the work of MPW}
In Ref.~\cite{MPW1}, MPW proved the absence of hard supersymmetry anomalies
in the on-shell component formalism.
They considered a general supersymmetric gauge theory whose lagrangian contains
dimensionless parameters only.
They proved that, except for a supersymmetric extension
of the ABJ anomaly, which we will assume to be absent, no other anomalies
occur.  The proof consists of two elements.  First they established
that, apart from the Wess--Zumino consistency conditions~(\ref{closed}),
several additional algebraic constraints may be imposed on
the breaking term $\db$.  They then showed that any solution of the complete
set of algebraic constraints is cohomologically
exact (with exception of the ABJ anomaly);
it can be removed by finite symmetry-restoring counterterms.

We now list these additional restrictions.
The constraints are first established for
the quantum part of the 1PI functional, $\G_q = \G_r - \Scl$,
and then extended to the breaking term $\db$.
Regarding the global BRST parameters,
it can be shown that $\G_q$ is independent of the translation and $R$-symmetry
parameters $\x_\m$ and $\c$.  Independence of $\x_\m$ is
automatically satisfied by $\db$ as well, whereas the independence of $\c$
can always be achieved by adding $R$-symmetry restoring counterterms.

As for the dependence on ghost-sector fields,
$\G_q$ and $\db$ are independent of the auxiliary
field $b$, and can depend on $\cbar$ only through the linear
combination $K_\m^A - \pa\m\cbar$,
where we recall that $K^A_\m$ is the source field coupled to the BRST
variation of the gauge field. Moreover, if we adopt the Landau gauge,
\ie, we take the limit $\a\to 0$ in Eq.~(\ref{Segf}), then
$\G_q$ and $\db$ can depend on $c$ only through its derivatives $\pa\m c$.

Let us briefly explain how these constraints arise.
The constraints satisfied by $\G_q$ can be shown to follow automatically
from elementary properties of the diagrammatic expansion.\footnote{
  The (finite) counterterms that will be introduced to eliminate
  the breaking term will be chosen to respect
  the same constraints as well. Since rigorous
  proofs have been given elsewhere (see Ref.~\cite{MPW1} and references therein),
  our emphasis is on explaining the physical origin of the constraints.
}
The terms in the classical action (\ie, in $S_{ext}$)
that depend on the BRST parameters $\x_\m$ and $\c$
are all linear in the dynamical fields.  Such terms cannot appear
as insertions in 1PI diagrams.  Therefore, $\G_q$ is independent
of $\x_\m$ and $\c$.  Similarly, $\G_q$ is independent of the auxiliary
field $b$, because $\Scl$ contains no interaction vertices that depend on $b$.

That the dependence of $\G_q$ on $\cbar$ can be only via the linear combination
$K_\m^A - \pa\m\cbar$ follows from the following observation.
Acting on the generating functional of 1PI diagrams with $\d/\d \cbar_a$
corresponds to making an insertion of the source term in the $c_a$ equation
of motion, namely, an insertion of $f_{abc} \pa\m (A_{\m b} c_c)$.  This,
in turn, is equal to $\pa\m$ acting on an insertion of the nonlinear part
of $s A_{\m a}$, to which the $K_{\m a}^A$ source field couples.

To constrain the dependence on $c_a$ we observe that
acting on a 1PI diagram with  $\d/\d c_a$ produces an insertion of
\begin{equation}
  f_{abc} (\pa\m \cbar_b) A_{\m c}
  = f_{abc} \pa\m(\cbar_b A_{\m c}) - f_{abc} \cbar_b \pa\m A_{\m c} \,.
\label{cbardA}
\end{equation}
The first term on the right-hand side
is a total derivative, and will generate dependence on
$\pa\m c_a$ only.  The last term involves the longitudinal part of
the gauge field, $\pa\m A_{\m c}$.
Having adopted the Landau gauge, the gauge-field
propagator is transversal, and thus an insertion of $\pa\m A_{\m c}$ into
any diagram vanishes.  The conclusion is that $\G_q$ can only
depend on $\pa\m c_a$.

We next turn to the
breaking term.  Because $\G_q$ is independent of $\x_\mu$ and of $\c$,
and since in itself the classical action satisfies the ST identity,
any dependence of the breaking term
$\db=\cs(\G_r)$ on $\x_\mu$ or $\c$ can only arise
from diagrams that violate the conservation of, respectively, momentum
and $R$ charge.  Momentum is conserved in virtually all regularization
methods, ruling out any dependence of $\db$ on $\x_\m$.

The $R$-charge conservation is often violated by the regularization
(\eg\ dimensional regularization).  Because $\G_q$ is independent of $\c$,
the $\c$-dependent terms in $\db$ originate from regularized
diagrams that violate the $R$ charge conservation,
whose sum collapses to a contact term when the regularization is removed.
When the ST operator is applied, the $R$-charge violating contact terms
present in $\G_r$ will give rise to $\c$-dependent terms in $\db$
(see Eq.~(\ref{Dbreaking})),
which, therefore, must take the form of $\c \sum_{q\ne 0} q O_q$,
where $q$, the $R$ charge of $O_q$, cannot vanish.
It follows that adding the $R$-symmetry restoring
counterterms $i\sum_{q\ne 0} O_q$
removes all the $\c$ dependent terms from $\db$, along with
any terms that have a nonzero $R$ charge, by removing all (contact)
terms with nonzero $R$ charge from $\G_q$.

Finally, a set of algebraic identities
analogous to the Wess--Zumino consistency conditions can be used
to show that the dependence of $\db$ on the ghost-sector fields
is subject to the same restrictions as $\G_q$ \cite{MPW1}.

Armed with the Wess--Zumino consistency conditions, and all the additional
constraints that must be satisfied by the breaking term,
MPW proved the following result.
In a supersymmetric gauge theory that contains dimensionless couplings only,
barring a supersymmetric extension of the ABJ anomaly,
the most general breaking term allowed by the complete set of algebraic
constraints is $\cb$-exact.  This implies that all the symmetries,
and supersymmetry in particular, can be simultaneously restored order by order.

In more detail, suppose we have adjusted to zero the breaking term
at all orders up to $n-1$.  MPW then establish that, at order $n$
(after removing the divergences via \eg\ minimal subtraction),
there exists some local operator $\cq^{(n)} = \int d^4x\, \cq^{(n)}(x)$
whose $\cb$ variation reproduces the breaking term:
$\db^{(n)} = \cb \cq^{(n)}$.  The ghost number of $\cq^{(n)}(x)$ is zero,
and its mass dimension is equal to four.  Since $\cb$ is nilpotent,
the breaking term is made to vanish
by any \textit{symmetry restoring} counterterm of the form
$-\cq^{(n)} + \cb X^{(n)}$.  Here $X^{(n)} = \int d^4x\, X^{(n)}(x)$
with $X^{(n)}(x)$ of dimension three and ghost-number $-1$.
It is to be fixed by imposing some renormalization conditions.

We conclude with a few technical comments.
First, our book-keeping is slightly
different from that of MPW in the following sense.
Our source action~(\ref{sext}) contains a term associated with the BRST
transformation rule of every ``dynamical field,'' \ie, every field
that is integrated over in the partition function.
In contrast, MPW \cite{MPW1} do not introduce source fields that couple to the
BRST variation of the $b$ and $\cbar$ fields, which is possible
because these BRST variations are linear in the dynamical fields.
Their definition of the ST operator accommodates this difference.
Denoting objects pertaining to Ref.~\cite{MPW1} with a prime,
we have $\Scl = \Scl' +  K^{\cbar} s\cbar + K^b sb$,
and likewise, $\G_r = \G'_r + K^{\cbar} s\cbar + K^b sb$ (see Eq.~(\ref{dKlin0})).
It is then straightforward to check that $\cs(\G_r) = \cs'(\G'_r)$.
In particular, our $\cs(\G_r)$ is always independent of $K^{\cbar}$ and $K^b$.
As for the linearized ST operator, our $\cb$ contains derivatives
with respect to $K^{\cbar}$ and $K^b$, whereas their $\cb'$ does not.
But since $\cs(\G_r)$ is independent of these source fields,
it follows that $\cb'\, \cs'(\G'_r) = \cb\, \cs(\G_r) = 0$ as well.
In conclusion, the slight difference in setting up the starting-point
classical theory has no effect on the quantum theory.

The absence of hard supersymmetry anomalies
was directly established by MPW only in the Landau gauge.
But since the supersymmetry current is a gauge invariant operator,
were a supersymmetry anomaly to exist for any nonzero value of the gauge
parameter $\a$ (or for a difference gauge condition), this would automatically
constitute a violation of gauge invariance too.  Since the only
known gauge anomaly is the ABJ anomaly
(which we assume to be absent throughout),
we expect, based on MPW's work, that
no hard supersymmetry anomaly should arise
for any other gauge parameter or gauge condition as well.

MPW's derivation was done using the on-shell formalism.
While one expects the off-shell formalism to have the same physical content,
no explicit proof of the absence of hard supersymmetry anomalies was given
for the off-shell formalism.  If we are to make use of MPW's result,
we are thus bound to use the on-shell formalism, too.

\section{\label{spur} Absence of soft supersymmetry anomalies}
In this section we turn to the subject of this paper,
which is the study of soft supersymmetry anomalies.
In effect, the presence of dimensionful parameters in the lagrangian
means that the anomaly could be a lower-dimension operator.
We begin in Sec.~\ref{photino} with the observation that,
in a supersymmetric gauge theory containing an abelian group,
indeed there are new,
cohomologically nontrivial, lower-dimensional candidate anomalies.

We will prove that, nevertheless,
these lower-dimension anomalies never occur.
Our derivation is based on the following construction.
Starting from the original, or ``target,'' theory,
we trade each mass parameter $M_{ij}$ of the superpotential~(\ref{W})
with a new, dynamical,
chiral multiplet controlled by a new Yukawa coupling.
The theory thus obtained is amenable to the analysis of MPW \cite{MPW1},
and can be renormalized to all orders while preserving all the classical
symmetries, including, in particular, supersymmetry.
The quantum target theory will be recovered from the extended theory
in the limit where all the
new Yukawa couplings are sent to zero.  The dynamical effects of the
new multiplets then vanish, while the original mass parameters
emerge as the vacuum expectation values (VEVs) of the ``frozen out''
scalar fields.
The quantum target theory reconstructed this way preserves all of
its classical symmetries to all orders, and is thus free of anomalies.

An outline of the construction is given in Sec.~\ref{outline},
and the rest of this section is devoted to its details.
A brief summary is given in Sec.~\ref{smr}.
Appendix~\ref{Omega} deals with the generalization to the case
that the superpotential contains dimension-two parameters $\O_i$.

\subsection{\label{photino} Abelian gaugino anomalies}
The breaking term will in general contain an $\h$ dependent part,
$\db_\h = \partial\db/\partial\h$, which is responsible for the
violations of supersymmetric Ward identities.
The mass dimension of $\db_\h$ is $9/2$.
In more detail, we may write
$\db_\h = \sum C_i \co_i$, where the coefficients $C_i$
are built from the parameters of the theory (and, possibly,
global BRST parameters), and the $\co_i$ are operators
made up of the dynamical fields and sources.

Apart from dimensionless Yukawa couplings,
the most general superpotential
depends on parameters $M_{ij}$ with mass dimension one, as
well as on parameters $\O_i$ with mass dimension two.
If these dimensionful parameters occur in one of the coefficients
$C_i$, the operator $\co_i$ that multiplies it has mass
dimension lower than $9/2$.  Thus, allowing for the most
general superpotential could give rise to new supersymmetry anomalies that
were not considered in Ref.~\cite{MPW1}.

\def\dbph{\db_2}

In a supersymmetric gauge theory that contains an abelian gauge field,
there are in fact new cohomologically nontrivial solutions.
Denoting by $\l(x)$ the abelian gaugino field, the super-partner
of the abelian $A_\m(x)$,  these new solutions are given by
$\dbph = \int d^4x\, \dbph(x)$, where
\begin{subequations}
\label{photinoan}
\begin{eqnarray}
  \dbph(x) &=& \cc\, \bh P_- \l(x) + \cc^*\, \bh P_+ \l(x) \,,
\label{photinoana}\\
  \cc &=& a_{ijklmn} M_{ij} M_{kl} M_{mn} +  b_{ijk} M_{ij} \O_k \,,
\label{photinoanb}
\end{eqnarray}
\end{subequations}
and where the coefficients $a_{ijklmn}$ and $b_{ijk}$ depend on
the dimensionless coupling constants only.  The subscript 2 is to remind
us of the mass dimension of the operator $\bh P_\pm \l$.\footnote{
  For our conventions, see Sec.~\ref{onsh}.
}

Let us prove that $\dbph$ is a cohomologically nontrivial
solution of the Wess--Zumino consistency conditions~(\ref{closed}),
\ie, that $\dbph$ is $\cb$-closed but not $\cb$-exact.\footnote{
  For further discussion of the divergence of the supersymmetry current
  in the presence of the anomaly~(\ref{photinoan}), see App.~\ref{MPW2spr}.
}

Closedness, $\cb \dbph =0$, is equivalent to verifying
that $\cb\dbph(x)$ is a total derivative.  Writing
$\cb = s + (\cb-s)$ we will show that $s\dbph(x)$ is a total derivative,
and $(\cb-s)\dbph(x)=0$.   To prove the former
we split up the on-shell BRST operator $s$ according to Eq.~(\ref{ssplit}),
and consider each term on the right-hand side.
The abelian gaugino field $\l(x)$ is gauge invariant, and so
$s_g \dbph(x) = 0$.  As for the supersymmetry part of the BRST transformation,
since $\h$ is commuting the (on-shell) $D$ term in the transformation
rule~(\ref{sonb}) drops out, and we have
$s_\h \, \bh P_\pm \l(x) \propto \bh P_\pm \s_{\m\n} F_{\m\n}(x) \h$.
This is a total derivative since the abelian field strength
is $F_{\m\n} = \pa\m A_\n - \pa\n A_\m$.\footnote{
  While closedness requires that $\cb\dbph(x) =\pa\mu \cj_\m$
  for some $\cj_\m$, the latter need not be a gauge invariant operator.
  Of course, $\dbph$ itself is gauge invariant.
}
Also, $s_\x \l(x)$ is trivially a total derivative.
Turning to the $R$-transformations part $s_\c$
we notice that, in this case, not just $\l(x)$ and $\h$
but also the parameters $M_{ij}$ and $\O_i$
transform nontrivially.
The $R$ charge assignments encoded in the transformation rules
of Sec.~\ref{onsh} imply that $\dbph$ has zero $R$ charge, hence
$s_\c \dbph=0$.  Finally, we have to take
into account the difference between $\cb$ and $s$ when
acting on $\l(x)$.
Contracting the right-hand side of Eq.~(\ref{Bsa}) with $\bh P_\pm$ yields zero.
This completes the demonstration that $\cb\dbph=0$,
\ie, that $\dbph$ is closed.

We next show that $\dbph$ is not exact, \ie, that there does not exist any
local operator $\cq(x)$ such that $\dbph = \cb\cq$,
with $\cq =\int d^4x\, \cq(x)$.  In a nut shell, the reason
is that $\bh P_\pm \l(x)$ is not the supersymmetry variation of anything.
From the simple structure of $\dbph$ it follows that if a $\cq$ existed,
then it would have to satisfy $s_\h \cq = \dbph$, as well as
$s_g \cq = s_\x \cq =s_\c \cq =(\cb-s) \cq = 0$ (\eg\ if it was not true
that $s_g \cq = 0$ then $\dbph$ would depend on the
ghost field $c(x)$).

The coefficients of $\bh P_\pm \l(x)$ have the generic form of ``$M^3$,''
for the first term on the right-hand side of Eq.~(\ref{photinoanb}),
and ``$M\O$'' for the second term.
Let us consider first the $M^3$ part. The mass parameters $M_{ij}$
do not occur alone in the transformation rule of any field
(only the combination $M_{ij} \f_j(x)$ occurs in the transformation rule
of $\j_i(x)$).  This means that $\cq(x)$ would have to have the form
$a_{ijklmn} M_{ij} M_{kl} M_{mn} X(x) +\hc,$ where $X(x)$ is linear in the fields,
has $R$ charge equal to $-2$, and satisfies $s_\h X(x) = \bh P_- \l(x)$
up to a total derivative.
Since no (dynamical or source) field has these properties,
the $M^3$ part of Eq.~(\ref{photinoan}) is not exact.

The situation for the $M\O$ term is slightly more involved,
since in this case we may obtain $\O_k\h$ from the supersymmetry variation
of $P_- \j_k(x)$.  This means that we have to consider the candidate
$\cq(x) = b_{ijk} M_{ij} \bj_k(x) P_-\l(x) +\hc$.  While the last term
on the right-hand side of Eq.~(\ref{photinoanb})
does occur in the $\cb$-variation
of this $\cq(x)$, there are of course additional terms in the variation.
In particular, it can be checked that the operator
$b_{ijk} M_{ij} \bj_k(x) P_- \s_{\m\n} F_{\m\n}(x) \h$,
that comes from applying $s_\h$ to $\l(x)$, does not occur in the variation
of any other operator except the above $\cq(x)$.
This proves that the $M\O$ term
is not exact, too.  The only difference is that,
in the case that $a_{ijklmn}=0$, it would be possible to
``maneuver'' the anomaly into other forms using the above
counterterm.  In particular, by tuning the coefficient
of that counterterm it would be possible
to construct representatives of the same cohomologically nontrivial solution
that do not contain any part linear in the abelian gaugino field.

The behavior of the candidate supersymmetry anomaly $\dbph$
under discrete symmetries is also important, because such symmetries, which
usually survive regularization, can therefore be used to constrain the
possibilities.  Under charge conjugation, the abelian gaugino field
flips sign, as does the abelian gauge field itself, whereas
the supersymmetry parameter $\h$ is invariant.  Therefore, $\bh P_\pm\l$
is charge-conjugation odd.  This rules out the existence of the
anomaly $\dbph$ in theories where charge conjugation is a good symmetry.

The Standard Model lacks charge conjugation symmetry.
It also has an abelian gauge field, the hypercharge field,
and a corresponding abelian gaugino.
The occurrence of the ``abelian gaugino'' anomaly $\dbph$
in supersymmetric extensions of the Standard Model
therefore cannot be ruled out by invoking charge conjugation.
We observe that also in supersymmetric gauge theories with a vector-like
matter content, charge conjugation could be explicitly broken
by the superpotential, in which case the occurrence of $\dbph$ would
again be allowed if an abelian gauge group is present.
As for parity, the linear combinations
$\bh \l$ and $\bh \g_5\l$ are parity-even and odd respectively,
and thus they are $CP$-odd and $CP$-even respectively.
A $CP$-even anomaly would not be suppressed by the smallness
of the $CP$-violating phase.

Both the $CP$-even and $CP$-odd parts of $\dbph$ could not arise at one loop.
The reason is that any diagram that contributes to them
would have to ``know'' both about
the absence of charge-conjugation symmetry, which, in the case of the Standard
Model, comes from the gauge interactions,
and about the dimensionful parameters themselves,
which originate from the superpotential.\footnote{
  Recall that we are only considering classically supersymmetric theories,
  and that a Higgs mechanism requires the existence of dimensionful
  parameters in the superpotential.
}
In addition, any flavor symmetry
that is broken by the massive part of the superpotential
would constrain the structure of the coefficients $a_{ijklmn}$ and $b_{ijk}$.

So far, we have seen nothing that would rule out the occurrence of these
candidate supersymmetry anomalies starting from some order in the loop expansion
in supersymmetric extensions of the Standard Model.
It is therefore imperative to determine whether they can actually show up,
or alternatively, to rule them out to all orders by some other reasoning.
As it turns out, a proof that these supersymmetry anomalies never occur
can be given.

\subsection{\label{outline} Outline }
In the rest of this paper we prove the absence of soft supersymmetry anomalies.
In field theory, masses are often related to the VEVs of scalar fields.
This motivates us to reconstruct our target theory from an extended
theory that contains additional scalar and fermion spurion fields,
and no dimensionful parameters.
In this extended theory, a special limit is then taken,
in which the new scalar fields are frozen out at chosen VEVs that recover
the target theory's dimensionful parameters.

This idea applies to all mass parameters with dimension one,
\ie, the parameters $M_{ij}$ in Eq.~(\ref{W}), in a rather natural way.  The
most general superpotential also contains parameters of dimension two,
the $\O_i$, and we will deal with these separately.  Any potential
soft supersymmetry anomaly can contain only one factor of $\O$, leading to an
operator of dimension three, or one factor of $\O$ and one factor of
$M$, leading to an operator of dimension two.
The list of potential anomalies with dimension three is rather restricted,
and we show that all such candidate anomalies
can be removed by counterterms in
App.~\ref{Omega}, by considering all candidates explicitly.  Those with
an extra factor of $M$ can then be excluded as an application of the
general theorem we develop in this section.\footnote{
  We leave aside the question whether also the case of soft candidate
  anomalies with a factor of $\O$ can be treated by spurion techniques.
}

We will thus trade each mass parameter $M_{ij}$ of the target theory with
a new chiral multiplet and a new Yukawa coupling.
This approach leads to a technical, but important, complication.
The on-shell supersymmetry transformation of the fermion member
of the new chiral multiplet will be non-linear,
and, furthermore, will depend on
the fields of the original theory.  We must ensure the order-by-order
locality of the breaking term under these circumstances.
The natural way to do that is to rely on the Regularized Action Principle
(see Sec.~\ref{ARrev}).  But since the Regularized Action Principle
is an expression of the quantum equations of motion,
this requires that
all fields in the extended theory be dynamical.

Our solution is thus to promote the spurions fields themselves from
external fields to new dynamical fields.  What we regain
is the locality of the breaking term, and
the applicability of MPW's theorem.
As a result, it will be possible to renormalize the extended theory
such that all the classical symmetries, including supersymmetry,
are restored order by order.

The price we pay is that the extended theory
now contains dynamical degrees of freedom not present in the target theory.
As a consequence,
there are new diagrams with internal ``dynamical spurion''
lines with no parallel in the target theory.
However, the unwanted new diagrams can be suppressed
in a natural way.  Each dynamical spurion multiplet comes with
a new Yukawa coupling constant.  This Yukawa coupling
controls its dynamical effects, and allows us to suppress those
by taking the limit where that coupling is sent to zero, while keeping
the corresponding mass parameter of the target theory fixed.

Establishing all-orders supersymmetry in the extended theory
does not necessarily imply the same for the target theory,
because the extended theory contains fields not present in the
target theory.  When the spurion fields are ``turned off,''
we must show that the counterterms that are needed to restore
supersymmetry at each order can be constructed using only
the original fields and parameters present in
the target theory.  The role of the extended theory will be only as a device,
used at intermediate stages, that ultimately allows us to prescribe
the symmetry-restoring counterterms within the target theory itself.

In the next subsection we introduce the ``dynamical spurion'' fields
and discuss the classical action of the extended, or spurionized, theory
in some detail.  The absence of hard supersymmetry anomalies
is used to restore all the symmetries
of the spurionized theory.  The new crucial element is despurionization,
whose precise definition will be given in the next subsection.
Loosely speaking, despurionization is the operation of ``turning off''
the spurion fields by sending the new Yukawa couplings to zero,
while holding the VEVs of the scalar spurion fields at the desired values.
We will show that, upon despurionization,
the counterterms of the spurionized theory reduce to a set of counterterms
that depend only on the fields and parameters of the target theory,
and that the renormalized target theory defined by these
counterterms preserves all the classical symmetries to all orders.

The rest of this section is devoted to the details of the proof.
In Sec.~\ref{main} we state the main theorem to be proven more precisely.
In Sec.~\ref{filt} we introduce the notion of filtration as it will be
used in this paper, and derive some technical results.
In Sec.~\ref{spurz} we discuss the renormalization of the spurionized
theory. Sec.~\ref{despur} deals with the construction of the quantum
target theory, and completes the main part of the proof.
Finally, in App.~\ref{Omega} we deal with the case that some $\O_i$
are nonzero,
while appendices~\ref{embed} and~\ref{wnct} provide proofs of two
technical lemmas.

\subsection{\label{dyntheory} Dynamical-spurion theory}
It is instructive to first work out a simple example.
Let us consider a supersymmetric theory with the superpotential
\begin{equation}
  W_m = m \f_+ \f_- + \cdots \,.
\label{Wmass}
\end{equation}
The fields $\f_\pm$ are the scalar members of on-shell chiral multiplets
$(\f_\pm,\j_\pm)$.  They belong to complex conjugate representations
of the gauge group.  The ellipses stand for unspecified additional
terms that in principle depend on all other (scalar) fields
of the target theory, and that may or may not depend on $\f_\pm$.
At this stage we assume that the rest of the superpotential
involves (dimensionless) Yukawa couplings only.
As an example, in supersymmetric QCD, the $\f_\pm$
can play the role of the two squarks associated with a given quark field,
all having the same mass $m$.

Spurionization proceeds by trading
the mass parameter $m$ with a new chiral multiplet $(\f_s,\j_s)$
and a new Yukawa coupling $w$.\footnote{
  Without loss of generality we may assume that $w$ is real.
}
The \textit{dynamical spurion} fields
$\f_s$ and $\j_s$ are gauge singlets.
The spurionized lagrangian is still given by Eq.~(\ref{onL}),
where now the matter content includes both the original fields of the target
theory, and the new spurion multiplet.  There are kinetic terms
for all fields, including the spurions.
Since the spurions are gauge singlets, their kinetic terms
contain no interaction terms, and there is no
Yukawa-gauge coupling between the spurions and any of the gauginos.
With the replacement $m \to w\f_s$ the spurionized superpotential becomes
\begin{equation}
  W_s = w \f_s \f_+ \f_- + \cdots \,.
\label{Wspur}
\end{equation}
This superpotential gives rise to new interactions that we will
discuss shortly.
The BRST transformation rules, given by the general formulae of Sec.~\ref{onsh},
now depend on the superpotential~(\ref{Wspur}).  Likewise, $S_{ext}$,
including both its $K$-linear and $K$-bilinear
terms, is constructed while treating the original dynamical fields,
and the new spurion fields, on equal footing.

The target theory is to be recovered by  \textit{despurionization}.
By definition, this is the process of taking the limit $w\to 0$,
while holding the VEV of the scalar spurion at
\begin{equation}
  \svev{w\f_s} = m \,.
\label{spurvev}
\end{equation}
For the fermionic partner we will have, of course, $\svev{\j_s}=0$.
Let us examine the effect of this operation at the classical level,
starting with $S_0$ (see Eq.~(\ref{scl})).  Despurionization
is facilitated by expanding the scalar spurion field as
\begin{equation}
  \f_s = m/w + \d\f_s \,,
\label{expandphis}
\end{equation}
with $\d\f_s$ being the fluctuating part, and then sending $w\to 0$.
The fluctuating part $\d\f_s$ decouples in this limit,\footnote{
  More precisely, the kinetic terms in the spurion sector turn into
  a decoupled, free, massless Wess--Zumino model.
}
and the classical action of the target theory is recovered from
that of the spurionized theory.

The spurionized lagrangian contains new interactions,
some of which are wanted, and some unwanted.
The wanted interactions are those that, under despurionization,
reduce to $m$-dependent terms of the target theory's lagrangian.
The unwanted interactions include all other $w$-dependent interactions.
In the purely bosonic sector the new interactions are
\begin{subequations}
\label{spurint}
\begin{eqnarray}
  | \partial W / \partial \f_s |^2 &=& |w\f_+\f_-|^2 \,,
\label{spurinta}\\
  | \partial W / \partial \f_\pm |^2 &=& |w\f_s\f_\mp + \cdots|^2 \,,
\label{spurintab}
\end{eqnarray}
\end{subequations}
where the ellipses in Eq.~(\ref{spurintab}) come from any
additional $\f_\pm$ dependent terms that may be present in the original
superpotential~(\ref{Wmass}).
Of the new interactions, $|w\f_s\f_\mp + \cdots|^2$ is
wanted, as it reduces to $|m\f_\mp + \cdots|^2$.
The other interaction, $|w\f_+\f_-|^2$, is unwanted,
since there is no matching term in the original lagrangian,
and indeed it vanishes in the limit $w\to 0$.
In the fermion sector, we have the wanted interaction
$w\bj_+ (\f_s P_+ + \f_s^* P_- ) \j_-$ that reduces to the mass term
$\bj_+ (m P_+ + m^* P_-) \j_-$ upon despurionization,
and the unwanted interactions $w\bj_s \f_\pm P_+ \j_\mp +\hc$,
that once again vanish in the limit $w\to 0$.

Turning to the source action $S_{ext}$, we need to specify what is to be done
with the source fields that couple to the BRST variations of the spurion-sector
fields. The fermionic spurion source  $\bK^{\j_s}$ is simply set to
zero.  For the scalar spurion's source $K^{\f_s}$,
we choose the despurionized value to be $(w/V)k^m$,
where $V$ is the space-time volume, and $k^m$ is the (global) source
for the BRST variation of the mass parameter $m$.\footnote{
  Similarly we set $K^{\f_s^*}=(w/V)k^{m^*}$.
}
Recall that the transformation rule for mass parameters, Eq.~(\ref{parone}),
means that we treat $m$ as a global spurion, because of $R$ symmetry.
The despurionization rule for $K^{\f_s}(x)$
has the effect that the source term for the
(local spurion) field $\f_s(x)$ reduces
to the source term for the (global spurion) parameter $m$.
Explicitly, using Eq.~(\ref{sonc}) we have
\begin{eqnarray}
  \int d^4x\, K^{\f_s}(x) s\f_s(x) &=&
  \int d^4x\, K^{\f_s}(x)
  \Big( \sqrt{2}\,\bh P_+\j_s(x) + \x_\n\pa{\n}\f_s(x)
  +i (2/3) \c \f_s(x) \Big)
\NON
    & \to & V (w k^m/V) i (2/3) \c\, m/w + \cdots
\NON
  &=& k^m i (2/3) \c m + \cdots \,,
\label{despKs}
\end{eqnarray}
where the arrow indicates the substitutions performed during
the despurionization process, and the
ellipses stand for terms that will vanish once we take the limit $w\to 0$.
Note that the factor of $1/w$ originating from the right-hand side
of Eq.~(\ref{expandphis}) is cancelled by the factor of $w$ coming
from the despurionization rule of $K^{\f_s}(x)$.

Looking ahead, our task will be to extend the scope of despurionization
consistently from the classical to the quantum theory.  In the remainder
of this subsection we touch upon some of the issues we will encounter.

The quantum spurionized theory may require symmetry-restoring counterterms
that depend on the fermionic spurion field $\j_s$,
whose supersymmetry variation contains the nonlinear terms
$w\f_+\f_- P_-\h$ and $w\f_+^*\f_-^* P_+\h$.
The worry is that these terms, which depend only on the fields of the
target theory, would be needed in
order to restore supersymmetry Ward identities of the target theory as well.
This would be a problem, because
in the target theory the spurion field $\j_s$ does not exist, and thus
no counterterm that depends on it can ever be constructed!

Taking the $w\to 0$ limit ensures that such an impasse will not arise.
Consider a Feynman diagram of the spurionized theory containing an
insertion of a nonlinear term coming from the $\j_s$ transformation
rule.  When we send $w\to 0$, all such diagrams vanish, for the simple
reason that an insertion of $w\f_+\f_-$ entails a factor of $w$.  What
this means is that $\j_s$-dependent counterterms may be needed to
restore supersymmetry Ward identities in the extended theory, but never in the
target theory.  As for $\f_s$-dependent counterterms, we will show that
they reproduce the $m$-dependent counterterms of the target theory upon
despurionization.\footnote{
  The supersymmetry variation of $\j_s$ also contains a
  term linear in $\pa\m \f_s$.  Since this term involves a derivative,
  the VEV of $\f_s$ drops out.  This is important, because the VEV of
  $\f_s$ is designed to survive despurionization
  (\seef~Eq.~(\ref{spurvev})), and thus it better not occur in the
  transformation rule of a field that does not survive despurionization.
}

Another subset of diagrams of the spurionized theory that vanishes
in the $w\to 0$ limit includes any diagram containing an
internal dynamical-spurion line.  The reason is that
spurion interactions always involve the coupling constant $w$.
Adding a dynamical-spurion propagator to a Feynman
diagram adds a factor of $w^2$.  All diagrams with a propagating spurion
line are thus suppressed in the limit $w\to 0$.
As a result, the promotion of the spurions from external
to dynamical fields does not prevent us from reconstructing
the target theory.  This, of course, is a key point of the whole
construction: we need spurions to be dynamical in order to
invoke MPW's theorem, but we then need to get rid of this
dynamics when we return to the target theory.

The example we have presented in detail above
contains a single mass parameter $m$.  The discussion
generalizes straightforwardly to a general $M_{ij}$.
The spurionization process amounts
to trading each nonzero entry $M_{ij}$ with a separate spurion multiplet,
along with its own Yukawa coupling.  MPW's theorem applies to the fully
spurionized theory, which, as before, contains no dimensionful parameters.
We may thus renormalize the fully spurionized theory to all orders while
preserving all of its symmetries.  We then ``descend'' back to the target
theory by applying the despurionization process
to all spurion multiplets.  The new Yukawa couplings are all sent
to zero, while the associated scalar-spurion VEVs are held
fixed at the desired values.   Since a general $M_{ij}$
does not lead to any new issues, and only requires a more elaborate
book-keeping, we will formulate the rest of this section in terms
of the example given above that contains the single dimensionful parameter $m$.

Similarly, the case of a candidate anomaly quadratic or cubic in the $M_{ij}$
is covered by our extension of the MPW theorem.  Our detailed study will
be concerned with the descent from dimension-five breaking terms (the
case to which MPW's theorem applies) to dimension-four breaking terms,
\ie, breaking terms linear in $M_{ij}$.
But, our theorem then also applies to the further descent to breaking
terms cubic or quadratic in the $M_{ij}$.
As already mentioned, an exception is
the case that an $\O_k$ appears in the breaking term.   This case is covered
by the analysis of App.~\ref{Omega}.

\subsection{\label{main} Statement of main result}
The spurionized theory has no dimensionful parameters.
By MPW's theorem \cite{MPW1}, its counterterm action can be chosen
such that the renormalized 1PI functional
satisfies the ST identity to all orders in perturbation theory,
\ie, such that the quantum spurionized theory preserves all
classical symmetries.

Our strategy is to use the spurionized theory as a device that helps
us prescribe how the target theory will be renormalized.
In a specific regularization scheme,
such as dimensional regularization, constructing the quantum target theory
amounts to prescribing the complete counterterm action order by order.
It goes without saying that, whatever role we envisage for the spurionized
theory, the counterterm action we select for the target theory
must depend only on the fields and the parameters present in that theory;
we have designed despurionization to effect this constraint.

Let us denote objects pertaining to the spurionized
theory by a check mark, while objects without a check mark will
belong to the target theory.  The main result of this paper is
the following theorem.

\mythrm{Theorem 1}{Reconstruction of the quantum target theory}
Let us fix the counterterm action of the target theory to be
that obtained by despurionization from the spurionized theory's one.
The following statements are then true order by order in perturbation theory:

\vskip 1ex\noindent {\bf 1(a)}
After the removal of infinities via minimal subtraction,
the quantum 1PI functional of the target theory $\G_q^{(n)}$
is the despurionized limit of the spurionized theory's one, $\ckG_q^{(n)}$.

\vskip 1ex\noindent {\bf 1(b)}
The breaking term of the target theory $\db^{(n)}$ may be obtained
by despurionization from the breaking term $\ckdb^{(n)}$
of the spurionized theory.

\vskip 1ex\noindent {\bf 1(c)}
Closedness of the breaking term in the spurionized theory,
$\ckcb \ckdb^{(n)} =0$,
implies the same in the target theory, $\cb \db^{(n)} =0$.

\vskip 1ex\noindent {\bf 1(d)}
The symmetry-restoring counterterm action $\ckS_{t,f}^{[n]}$
of the spurionized theory, which satisfies $\ckdb^{(n)} = -\ckcb \ckS_{t,f}^{[n]}$,
can be chosen such that it
reduces upon despurionization to a symmetry-restoring counterterm action
for the target theory $S_{t,f}^{[n]}$
that satisfies $\db^{(n)} = -\cb S_{t,f}^{[n]}$.

\vskip 1ex\noindent
The outcome is an all-orders renormalized target
theory that preserves all of its classical symmetries also at the quantum level.
The technical ingredients needed for the proof of Theorem~1 will be gradually
developed in the following subsections.  The proof itself,
which puts together all these ingredients, is given in Sec.~\ref{despur}.

Before delving into the technical details let us make a few comments.
First, it is intuitively clear that the operation of removing the
infinities via minimal subtraction commutes with despurionization.
A detailed justification of this observation will be given below.

We will renormalize the spurionized theory by developing the perturbative
expansion around the classical vacuum where all scalar VEVs are zero.
This is a supersymmetric minimum of the classical potential.
The spurionized theory is massless on the chosen vacuum, and,
by construction, free of any dimensionful parameters.
We may thus impose renormalization conditions at some nonzero but otherwise
arbitrary $p^2=\m^2$ \cite{BM}.

With our definition of the quantum target theory, the despurionization process
extends straightforwardly to individual Feynman diagrams.
To this end, we simply have to develop perturbation theory
in the spurionized theory around a different classical vacuum: the one
specified by Eq.~(\ref{spurvev}).
This involves splitting the scalar spurion field into a classical
and a fluctuating part according to Eq.~(\ref{expandphis}).\footnote{
  It is not required that the classical vacuum with $\svev{\f_s}=m/w$ will be
  a (local) minimum of the classical potential.  Adding the usual source terms
  $\sum_I \int d^4x\, J_I(x) \F_I(x)$ to the generating functional
  for connected diagrams, we enforce the desired $\f_s$ VEV by adjusting
  the (constant mode of the) corresponding source $J_s$ to the needed,
  in general nonzero, value.
}
We will keep using precisely the same set of
counterterms that were previously determined
on the massless vacuum.  Here we are making the standard assumption that,
being the divergence of a Noether current, an anomaly is a local operator.
As such, it is independent of the particular vacuum state chosen to develop
perturbation theory.  It follows that those (symmetry restoring) counterterms
that eliminate the breaking term, and restore all the symmetries
of the spurionized theory on the massless vacuum,
will do so on any other vacuum state as well.

We will not need to explicitly specify the renormalization conditions.
However, certain choices of the symmetry-restoring
counterterms of the spurionized theory would hamper the despurionization
process, and must therefore be avoided; that the (renormalization conditions
and the) symmetry-restoring counterterms can be chosen to comply with
this requirement will be proved later on.

The target theory may in general contain massless particles,
and may in general contain super-renormalizable couplings.\footnote{
  When more than one spurion multiplet is needed, a similar statement
  applies to the theories encountered at intermediate steps
  of the despurionization process.
}
This is the generic situation where infrared divergences could possibly
interfere with the renormalization process.
We will assume that our target theory is well-defined in the
infrared, so that this situation is avoided.
For a discussion of this issue in the context of supersymmetric extensions of
the Standard Model, see Ref.~\cite{ARrev}.

\subsection{\label{filt} Filtrations}
Let us use the generic name $B$ for any of the nilpotent operators
of interest in this paper.  The cohomology space of $B$ is constructed from
the space of all integrated local operators $X=\int d^4x\, X(x)$
that satisfy closedness, $BX=0$.
Two elements $X_1$ and $X_2$ in that space belong to the same equivalence class
if their difference is $B$-exact: $X_1-X_2 = B Q$, for
some $Q=\int d^4x\,Q(x)$. The cohomology space
is defined as the space of all equivalence classes.
The cohomology space is a linear vector space, in which the
zero vector is
the zero equivalence class consisting of all $B$-exact elements.
We will often refer to any element which is not $B$-exact as $B$ nontrivial.

Without any further restrictions, the complete cohomology space
will typically be infinite dimensional.  Finite-dimensional cohomology
spaces can be defined by prescribing a set of quantum numbers.
A quantum-number operator, denoted $\cq$,
will be by definition any operator for which $[\cq,B] \propto B$.
In general both fields and parameters may have nonzero quantum numbers;
concrete examples of quantum numbers will be encountered below.
The commutation relation above means that $B$ moves between spaces
with definite quantum numbers, and that the full cohomology space
may be divided into subspaces with definite quantum numbers.

Filtrations are a standard part of the mathematician's toolkit
for dealing with cohomology spaces.  Like a quantum number operator,
a filtration $\cn$ assigns a certain charge to every field and parameter.
The basic difference is that $B$ does not
satisfy any particular commutation relations with $\cn$.
The filtrations we will encounter have in common that $B$ splits
into terms that either maintain or raise, but never lower,
the filtration number $\cn$.  In other words, we have
\begin{equation}
  B = B_{\und{0}} + \cdots + B_{\und{m}},
\label{splitB}
\end{equation}
where $B_{\und{n}}$ denotes that part of $B$ that raises the filtration
number $\cn$ by $n$ units: $[\,\cn,B_{\und{n}}\,]=n B_{\und{n}}$.
An immediate consequence of these
definitions is that $B_{\und{0}}$, the part of $B$ that maintains
the filtration number, must be nilpotent too.  To see this,
we substitute Eq.~(\ref{splitB}) into the nilpotency relation satisfied by $B$,
namely $B^2=0$.   Each part of $B^2$ that raises $\cn$ by some fixed
amount must vanish separately.  In particular, the vanishing of the part
that does not raise $\cn$ gives rise to $B_{\und{0}}^2=0$.

\begin{table}
\vspace*{3ex}
\begin{center}
\begin{tabular}{ c | l } \hline \hline
$\cql$ & objects  \\ \hline \hline
+1      & $Y_{ijk}$, all gauge couplings \\ \hline
0       & $M_{ij}$, $\h$, $\x_\m$, $\c$. \\ \hline
-1      & $\O_i$, all fields   \\ \hline\hline
\end{tabular}
\end{center}
\begin{quotation}
\caption{Loop-number assignments, referring to the general
  superpotential~(\ref{W}).  All dimensionless coupling constants
  have $\cql=1$.  All fields have $\cql=-1$.
  This includes the dynamical fields, the corresponding effective fields
  on which the 1PI functional depends, and the
  $K$ source fields.  The global BRST parameters have $\cql=0$.
\label{tabloop}}
\end{quotation}
\vspace*{-4ex}
\end{table}

We next prove a lemma on the relation between
cohomology spaces induced by a filtration.

\mythrm{Lemma 2}{Filtration as embedding of cohomologies}
Consider a nilpotent operator $B$ and a filtration $\cn$
such that $B = B_{\und{0}} + \cdots + B_{\und{m}}$ where $m<\infty$.
Assume also that the cohomology space of $B$ with given quantum numbers
is finite dimensional.
For an element $X$ of the cohomology space,
the filtration is $X = X_{\und{k}} + X_{\und{k+1}} + \cdots + X_{\und{l}}$,
where $\cn X_{\und{n}} = n X_{\und{n}}$.
To set the conventions unambiguously, we assume
that the parts of $X$ with filtration number $\cn<k$ or $\cn>l$ vanish.
In addition, there exist $n_{min}$ and $n_{max}$ that depend only
on the quantum numbers of the cohomology space, such that
$-\infty < n_{min}\le k\le l\le n_{max} < \infty$.  Under these assumptions,
if $X$ is $B$ nontrivial, then it has a representative $X' = X + BQ$
whose lowest filtration part $X'_{\und{k'}}$
has $k\le k' \le n_{max}$, and is $B_{\und{0}}$ nontrivial.

\vskip 1ex

Although Lemma~2 is a standard result (see \eg\ Ref.~\cite{PgSr})
we give its proof in App.~\ref{embed}.
The reason is that we will be applying it
to a filtration with somewhat unusual properties, and thus it is
important to see precisely how the assumptions of the lemma enter the proof.

We now introduce
the specific quantum number operators and filtrations that we will need.
We begin with the quantum numbers.  Two of them were already
encountered in Sec.~\ref{onsh}.  These are the ghost number $\cqgh$,
and the operator that counts the mass dimension, $\cq_d$.
Referring to the linearized ST operator of the spurionized theory,
the action of $\ckcb$ increase $\cqgh$ by one, and, with our conventions,
the same is true for the mass dimension $\cq_d$.  (As will be seen shortly,
the other nilpotent operators of interest are contained in $\ckcb$,
and thus they change the quantum numbers by the same amount.)

A third quantum number that will play a role in our discussion
has its obvious origin in the diagrammatic expansion.
It is the loop(-counting) number $\cql$.
The $\cql$ assignments are given in Table~\ref{tabloop}, and
they are designed such that the classical action has $\cql=-2$,
while $n$-loop terms in the 1PI functional have $\cql=-2+2n$.
It can be checked that $\cql$ commutes with $\cb$.

\begin{table}
\vspace*{3ex}
\begin{center}
\begin{tabular}{ c | l } \hline \hline
$\cnw$ & objects  \\ \hline \hline
+1      & $w$ \\ \hline
-1      & $\f_s,\ \f_s^*,\ \j_s,\ K^{\f_s},\ K^{\f_s^*},\ \bK^{\j_s},$  \\ \hline
0       & everything else \\ \hline\hline
\end{tabular}
\end{center}
\begin{quotation}
\caption{$w$-number assignments.
\label{tabwn}}
\end{quotation}
\vspace*{-4ex}
\end{table}

We next introduce two filtrations that are
both related to the Yukawa coupling $w$ introduced in Sec.~\ref{dyntheory}
to control the spurion sector.
The first one will be called the $w$-number filtration, and it is defined
by the assignments of $w$-number, or $\cnw$, given in Table~\ref{tabwn}.
A glance at the table reveals that
the target theory contains no objects with nonzero $w$-number.
This tells us that the target theory will have to be recovered
from the $\cnw=0$ sector of the spurionized theory.
The $\cnw>0$ sector of the spurionized theory will vanish
when we apply the despurionization process,
because of the involved $w\to 0$ limit.
The spurionized action also contains terms with negative $w$-number,
but, as we will see, they occur only at the classical level, and do not disrupt
the construction of the quantum target theory.

Under the $w$-number filtration, the decomposition of the classical
action and the linearized ST operator of the spurionized theory is\footnote{
  We drop the check mark from objects with an underlined subscript
  since that subscript automatically identifies them as belonging
  to the spurionized theory.
}
\begin{subequations}
\label{spurwn}
\begin{eqnarray}
  \check{S}_{cl} &=& (\Scl)_{\und{-2}} +(\Scl)_{\und{0}} +(\Scl)_{\und{2}} \,,
\label{spurwna}\\
  \ckcb &=& \cb_{\und{0}} + \cb_{\und{2}} \,.
\label{spurwnb}
\end{eqnarray}
\end{subequations}
In the classical action,
the bilinear terms that depend on spurion-sector fields have $\cnw=-2$.
This includes the kinetic terms, the terms in $S_{ext}$ that account
for the linear part of the spurion-field transformation rules,
and the term bilinear in $\bK^{\j_s}$.   As for the remaining terms,
it is easy to keep track of the $w$-number if we remember that
the superpotential has $\cnw=0$.  Most other terms in
the classical action have $\cnw=0$.  The only term with $\cnw=2$
is shown in Eq.~(\ref{spurinta}).

As for the linearized ST operator $\ckcb$,
it has a part $\cb_{\und{0}}$ that does not change $\cnw$,
and a part $\cb_{\und{2}}$ that increases $\cnw$ by two units.
Terms that contribute to $\cb_{\und{2}}$ occur in the transformation rule
of $\j_s$;  these are the $w\f_+\f_- P_-\h$ and $w\f_+^*\f_-^* P_+\h$ terms
mentioned already.  Such terms also occur in the equation of motion part
(see Eq.~(\ref{Btrans})) of the transformation rules of some
of the $K$-sources.

The second filtration is defined by simply counting powers of $w$ itself.
It assigns a unit filtration number to $w$, and zero to everything else.
While just being a sophisticated name for the Taylor series in $w$,
it will nevertheless be useful to think about
the Taylor series as a filtration.
Using an overlined superscript to label the power of $w$,
we then have the following expansions
\begin{subequations}
\label{wonly}
\begin{eqnarray}
  \cb_{\und{0}} &=& \cb_{\und{0}}^{\ov{0}} + w \cb_{\und{0}}^{\ov{1}}
   + w^2 \cb_{\und{0}}^{\ov{2}} \,,
\label{wonlya}\\
  \cb_{\und{2}} &=& w \cb_{\und{2}}^{\ov{1}} + w^2 \cb_{\und{2}}^{\ov{2}} \,.
\label{wonlyb}
\end{eqnarray}
\end{subequations}
An immediate consequence is that $\cb_{\und{0}}^{\ov{0}}=\ckcb\Big|_{w=0}$.
In words, $\cb_{\und{0}}^{\ov{0}}$ is just the linearized ST operator
of the spurionic theory for the special case that $w=0$.

Our main interest will be in the breaking-term cohomology spaces
of the three nilpotent operators $\ckcb$, $\cb_{\und{0}}$
and $\cb_{\und{0}}^{\ov{0}}$.  We recall that a breaking-term space
is the cohomology space of closed integrated local operators
with ghost-number $\cqgh=1$ and mass dimension $\cq_d=1$.
We now argue that the breaking-term space of all three operators
is cohomologically trivial.\footnote{
  Recall we are assuming the absence of the ABJ anomaly.
}

To begin with, MPW's theorem directly applies to $\ckcb$,
which is the linearized ST operator associated with the classical action
of the spurionized theory.  Hence its breaking-term cohomology is trivial.
Next, according to the observation we have
made below Eq.~(\ref{wonly}), the breaking-term cohomology
of $\cb_{\und{0}}^{\ov{0}}$ is trivial, too, because $\cb_{\und{0}}^{\ov{0}}$
is merely the linearized ST operator derived from the spurionized
classical action in the special case that $w=0$.

Establishing the same result for the breaking-term space
of $\cb_{\und{0}}$ requires some work, because unlike the previous two
cases, $\cb_{\und{0}}$ is not the linearized ST operator associated with
any classical action.  Having ``bracketed'' $\cb_{\und{0}}$
between $\ckcb$ and $\cb_{\und{0}}^{\ov{0}}$, both of which are directly
under the scope of MPW's theorem, we prove the same result for $\cb_{\und{0}}$
using the embedding lemma, Lemma~2.

\mythrm{Lemma 3}{
The breaking-term cohomology of $\cb_{\und{0}}$ is trivial}

\vskip 1ex

We will prove this statement by applying Lemma~2 to the
powers-of-$w$ filtration.  Before we can do that, we must make sure
that the conditions of the lemma are satisfied.
In particular, the powers of $w$ that we may encounter
must belong to a bounded range.  This requirement is tricky,
because $w$ itself does not transform under $\cb_{\und{0}}$.
Therefore, a $\cb_{\und{0}}$-closed element $X_{\und{k}}$
(\ie, a solution of the equation $\cb_{\und{0}} X_{\und{k}} =0$
with $\cnw=k$) remains $\cb_{\und{0}}$-closed
if we multiply it by an arbitrary power of $w$.
The solution is to invoke the loop-counting number $\cq_l$.
At order $n$ in perturbation theory the breaking term $\db^{(n)}$
must have $\cq_l = -2+2n$.  Prescribing $\cq_l$ on top of $\cqgh$ and $\cq_d$
ensures that arbitrarily large powers of $w$ cannot occur,\footnote{
  Negative powers of $w$ never occur in the diagrammatic expansion.
}
as well as that the breaking-term cohomology space of $\cb_{\und{0}}$
is finite dimensional.

In order to prove Lemma~3, we now assume on the contrary that
the breaking-term cohomology of $\cb_{\und{0}}$ includes a nontrivial
element $X_{\und{k}}$.  Applying Lemma~2
it follows that $X_{\und{k}}$ has a representative
whose leading term in the Taylor series in $w$ is
$\cb_{\und{0}}^{\ov{0}}$ nontrivial.  This, however, is impossible,
because MPW's theorem applies to $\cb_{\und{0}}^{\ov{0}}$,
and its breaking-term cohomology is trivial.

\vskip 1ex

In the next subsection we will use the triviality of the
breaking-term cohomology of $\cb_{\und{0}}$ to construct
a set of symmetry-restoring
counterterms for the spurionized theory that, after despurionization,
will be adequate for the target theory as well.

\subsection{\label{spurz} Renormalization of the spurionized theory}
After the preparatory steps of the previous subsection
we now turn to a key feature of the spurionized theory.

\mythrm{Theorem 4}{Avoiding $\cnw<0$ counterterms}
In the spurionized theory the following is true order by order:

\vskip 1ex\noindent {\bf 4(a)}
The minimal subtraction counterterms have $\cnw\ge 0$.

\vskip 1ex\noindent {\bf 4(b)}
The breaking term has $\cnw\ge 0$.

\vskip 1ex\noindent {\bf 4(c)}
The symmetry-restoring counterterms can be chosen to have $\cnw\ge 0$.

\vskip 1ex\noindent
What Theorem~4 tells us is that the complete counterterm action
of the spurionized theory can be chosen
to have $\cnw\ge 0$.  Likewise, the renormalized quantum 1PI functional
$\ckG_q^{(n)}$ will have $\cnw\ge 0$.

As a preliminary step let us consider the regularized 1PI functional
before the introduction of any counterterms.  Disregarding its tree-level part
it is easy to see that this functional has $\cnw\ge 0$.
Indeed the only fields that carry negative $w$-number are
the spurion sector fields (see Table~\ref{tabwn}).  But $K^{\f_s}$ cannot occur
in loop diagrams since there is no interaction vertex that depends on it.
The remaining spurion-sector fields $\f_s$, $\j_s$ and $\bK^{\j_s}$
always occur in the interaction lagrangian multiplied by $w$.
Hence, before the introduction of any counterterms, the quantum
part of the 1PI functional can be expressed as a functional of
$w\f_s$, $w\j_s$ and $w\bK^{\j_s}$, and positive powers of $w$.

Turning to the inductive proof, we assume that Theorem~4 is
true up to order $n-1$.  An insertion of a counterterm
with $\cnw\ge 0$ into any diagram can either maintain or raise, but not
lower, the $w$-number of the diagram.  Therefore, before any $n$-th order
counterterms are included, the $O(\hbar^n)$ regularized 1PI diagrams,
both with and without lower-order counterterm insertions, all have $\cnw\ge 0$.

The first counterterms that we introduce are
the $n$-th order minimal subtraction counterterms, which
remove the overall divergences of the regularized $O(\hbar^n)$ diagrams.
Evidently, a minimal-subtraction counterterm has the same $w$-number
as the (sum of) diagrams whose divergence is being subtracted.
It follows that the $n$-th order minimal subtraction counterterms have
$\cnw\ge 0$.  The same is true for the $O(\hbar^n)$ quantum 1PI functional,
$\ckG_q^{(n)}=\ckG_r^{(n)}-\ckS_0$, obtained at this intermediate step
in the limit in which the regulator is removed (in dimensional
regularization, the limit
$d\to 4$).

We next show that after minimal subtraction, and before the introduction
of any symmetry-restoring counterterms, the breaking term $\ckdb^{(n)}$
has $\cnw\ge 0$.  We have established that at this point the only part
of $\ckG_r^{(n)}$ with negative $w$-number consists of the $\cnw=-2$
spurion-dependent bilinear terms in the classical action $\ckS_0$.
Let us now substitute the definition of the ST operator~(\ref{STop})
into that of the breaking term~(\ref{Dbreaking}), and consider separately
the terms that involve differentiation with respect to spurion-sector
fields, and the rest.  In the latter case,
when we differentiate with respect to any other field or parameter,
the $\cnw=-2$ part of $\ckS_0$ drops out.
The rest of $\ckG_r^{(n)}$ has $\cnw\ge 0$, and the same will be true
for the contribution to the breaking term.

As for the terms that arise from
differentiation with respect to spurion-sector fields,
the most dangerous ones are $(\d \ckS_0/\d\f_I)(\d \ckG_q^{(n)}/\d K_I)$
and $(\d \ckG_q^{(n)}/\d\f_I)(\d \ckS_0/\d K_I)$,
where here $\f_I$ stands for $\f_s$, $\f_s^*$ or $\j_s$, and
$K_I$ is the corresponding source field.\footnote{
  The classical terms $(\d \ckS_0/\d\f_I)(\d \ckS_0/\d K_I)$ vanish
  when summed over all fields and parameters
  because the classical action satisfies the ST identity.
}
Differentiation with respect to a spurion-sector
field raises $\cnw$ by one unit, and so
$\d \ckS_0/\d\f_I$ and $\d \ckS_0/\d K_I$ have $\cnw\ge -1$,
whereas $\d \ckG_q^{(n)}/\d \f_I$ and $\d \ckG_q^{(n)}/\d K_I$
have $\cnw\ge 1$.  It follows that both
$(\d \ckS_0/\d\f_I)(\d \ckG_q^{(n)}/\d K_I)$ and
$(\d \ckG_q^{(n)}/\d\f_I)(\d \ckS_0/\d K_I)$ have $\cnw\ge 0$.
By a similar reasoning, the remaining contribution
$(\d \ckG_q^{(n)}/\d\f_I)(\d \ckG_q^{(n)}/\d K_I)$ has $\cnw\ge 2$.
This completes the proof that $\ckdb^{(n)}$ has $\cnw\ge 0$.

The last step is to show that, given a breaking term with $\cnw\ge 0$,
the symmetry restoring counterterms can be chosen
to have $\cnw\ge 0$ as well.  The proof, which rests on Lemma~3,
is given in App.~\ref{wnct}.

The outcome is that the singular and the finite $n$-th order counterterms
all have $\cnw\ge 0$,
and thus the same is true for the renormalized, quantum 1PI functional
$\ckG_q^{(n)}$ at this order.  This completes the inductive proof.

\vskip 1ex
As a corollary, we may establish the statement, mentioned already
in Sec.~\ref{dyntheory}, that any renormalized 1PI diagram
with an internal spurion-sector line must have $\cnw \ge 2$.
The vertices at which the two ends of the (dynamical-)spurion propagator
are attached give rise to a factor of $w^2$ (with \textit{no}
compensating factors of the effective fields $\f_s$ or $\j_s$).
In addition, from Theorem~4 we know that all other interaction vertices
and counterterm insertions cannot lower the $w$-number of the diagram.

\subsection{\label{despur} Reconstruction of the quantum target theory}
As announced in Sec.~\ref{main},
the target theory's counterterm action is obtained by despurionization.
In the previous subsection we have established that the counterterm
action of the spurionized theory can be chosen to have $\cnw\ge 0$.
This is the prerequisite that will allow us to prove Theorem~1.

Let us begin with Theorem~1(a).  Since the quantum 1PI functional
$\ckG_q$ of the spurionized theory has $\cnw\ge 0$, and since
diagrams with internal spurion lines have $\cnw\ge 2$, it is clear that
the (diagrammatic expansion of the) target theory emerges from
the $\cnw=0$ sector of the spurionized theory.  Also, the minimal-subtraction
counterterms of the spurionized theory obviously remove its infinities for
any set of values of the coupling constants,
including, in particular, $w$.
Therefore the same
is true after despurionization, \ie, the minimally-subtracted 1PI functionals
satisfy
\begin{equation}
  \ckG_q |_{despur} = \G_q \,.
\label{despurG}
\end{equation}
Generating functionals of 1PI diagrams
depend on effective fields, to which any desired value can be assigned.
Accordingly, $(\cdot)|_{despur}$ means that the spurion sector's
effective fields are set to $\f_s=m/w$ for the scalar, and $\j_s=0$
for the fermion.\footnote{
  The values assigned to the spurion sector's source fields are those
  specified in Sec.~\ref{dyntheory}.
}

Proving  the claims of Theorem~1(b) through 1(d) requires us
to show that the algebraic steps needed for the successful construction
of the symmetry-restoring counterterm action commute with despurionization.
In more detail, Theorem~1(b) will follow by showing that despurionization
commutes with the action of the ST operator, whereas Theorem~1(c,d)
will follow by showing that despurionization commutes with the action of
the linearized ST operator.

We consider Theorem~1(b) first.
At each order, the breaking term is defined to be the result
of acting with the ST operator on the minimally-subtracted 1PI functional.
Therefore, Theorem~1(b) amounts to the following equation:\footnote{
  The breaking term of the spurionized theory is local (Sec.~\ref{ARrev}),
  and, by Eq.~(\ref{commST}), the same is true for the target theory.
}
\begin{equation}
  \cs(\ckG_r) \Big|_{despur} =  \cs ( \G_r  )\,.
\label{commST}
\end{equation}
As in the case of Theorem~4
we will prove Eq.~(\ref{commST}) by substituting the definition
of the ST operator~(\ref{STop}) into the definition of the
breaking term~(\ref{Dbreaking}), and considering the various contributions
one by one.\footnote{
  As noted in the previous subsection,
  the classical action always satisfies the ST identity,
  and thus the purely classical terms vanish on both sides of Eq.~(\ref{commST})
  as well.
}

When $\f_I$ is (an effective field associated with)
a dynamical field of the target theory, we must show that
\begin{subequations}
\label{dbspur}
\begin{equation}
  \frac{\d\ckG_r}{\d\f_I(x)} \frac{\d\ckG_r}{\d K_I(x)} \bigg|_{despur}
  =
  \frac{\d\G_r}{\d\f_I(x)} \frac{\d\G_r}{\d K_I(x)} \,.
\label{dbspura}
\end{equation}
Because Eq.~(\ref{dbspura}) does not involve differentiation with respect
to any spurion-sector field, its validity follows immediately from
the definition of despurionization and Theorem~4.
A similar result holds for the terms associated with the
transformation rules of the global BRST parameters $\x_\m$, $\h$ and $\c$.

As for the spurion-sector fields, we must show that
\begin{eqnarray}
  \int d^4x \,
  \frac{\d\ckG_r}{\d\f_s(x)} \frac{\d\ckG_r}{\d K^{\f_s}(x)} \bigg|_{despur}
  &=&
  \frac{\d\G_r}{\d m} \frac{\d\G_r}{\d k^m}
  \;=\;
  i(2/3)\c m\, \frac{\d\G_r}{\d m} \,,
\label{dbspurb}\\
  \frac{\d\ckG_r}{\d\j_s(x)} \frac{\d\ckG_r}{\d \bK^{\j_s}(x)} \bigg|_{despur}
  \rule{0ex}{4ex}  &=& 0 \,.
\label{dbspurc}
\end{eqnarray}
\end{subequations}
Considering Eq.~(\ref{dbspurc}) first, we break up the left-hand side
into four terms by writing $\ckG_r = \ckS_0 + \ckG_q$.
The functional derivatives of the classical action are given explicitly by
\begin{subequations}
\label{dS0djs}
\begin{eqnarray}
  \frac{\d \ckS_0}{\d \bK^{\j_s}}
  &=&
  \sqrt{2} P_+ \Big( \sl{\partial} \f_s - w\f_+^*\f_-^* \Big)\h
  + \sqrt{2} P_- \Big( \sl{\partial} \f_s^* - w\f_+\f_- \Big)\h
\label{dS0djsa}\\
  &&  + \x_\m \pa\m \j_s -i(1/3)\c\g_5\j_s
  +2 P_+ \h\, (\bK^{\j_s} P_- \h) +2 P_- \h\, (\bK^{\j_s} P_+ \h)\,,
\NON
  \frac{\d \ckS_0}{\d \j_s}
  &=&
  \pa\m \bj_s \g_\m - w (\bj_-\f_+ + \bj_+\f_-) P_+
  - w (\bj_-\f_+^* + \bj_+\f_-^*) P_-
\label{dS0djsb}\\
  && - \sqrt{2} \bh (K^{\f_s} P_+ + K^{\f_s^*} P_-)
  + \x_\m \pa\m \bK^{\j_s} + i(1/3) \c \bK^{\j_s} \g_5 \,.
\nonumber
\end{eqnarray}
\end{subequations}
Once the prescribed values are assigned to the (effective and source)
spurion sector fields, the right-hand sides become $O(w)$,
and so they vanish for $w\to 0$.
We stress that these values are assigned only after having performed
the functional differentiations indicated on the left-hand sides.
It follows that the terms
that involve $\d \ckS_0/\d \bK^{\j_s}$ and/or $\d \ckS_0/\d \j_s$
on the left-hand side of Eq.~(\ref{dbspurc}) vanish.
In addition, because each differentiation with respect
to a spurion-sector field adds one unit of $w$-number,
the purely quantum term
$(\d \ckG_q^{(n)}/\d\j_s)(\d \ckG_q^{(n)}/\d \bK^{\j_s})$
has $\cnw\ge 2$, and so it vanishes upon despurionization as well.

It remains to prove Eq.~(\ref{dbspurb}).
The quantum part $\ckG_q$ is independent of $K^{\f_s}$, whereas
$\d \ckS_0/\d K^{\f_s} = \check{s} \f_s$ produces the (unrenormalized)
BRST variation of $\f_s$.
By considerations similar to those made in Eq.~(\ref{despKs})
(in particular, regarding the $w$ dependence of the various factors)
it is now straightforward to establish Eq.~(\ref{dbspurb}): varying
the spurionized 1PI functional with respect to the local spurion field
$\f_s$, and then despurionizing, produces the same result as
first despurionizing, and then varying the global-spurion parameter $m$.
This completes the proof of Eq.~(\ref{commST}), and thus, of Theorem~1(b).

The final task is to prove that despurionization commutes with the linearized
ST operator.  The spaces of (integrated) local operators on which we must
demonstrate this commutativity correspond to the breaking term
and to the (symmetry-restoring) counterterm action.
In the spurionized theory, Theorem~4 tells us that these spaces are
constrained by $\cnw\ge 0$.  Any operator with $\cnw\ge 0$
can be expressed as $\ckco=\ckco(w\f_s,w\j_s,w\bK^{\j_s};w)$.\footnote{
  Recall that both $\ckG_q$ and $\ckdb$ are independent of $K^{\f_s}$.
}
This notation means that $\ckco$ depends on the spurion-sector fields
only through $w\f_s$ \etc, and that, on top of that,
any explicit $w$ dependence must be polynomial.
Of course, $\ckco$ is allowed to depend on all
the fields, the $K$ sources, and the dimensionless parameters
of the target theory as well (this dependence will be suppressed).
Given any operator $\ckco$ with these properties, we have
\begin{subequations}
\label{BBO}
\begin{equation}
  \Big(
  \ckcb\, \ckco(w\f_s,w\j_s,w\bK^{\j_s};w)
  \Big) \Big|_{despur} = \cb\, \co(m)\,,
\label{BBOa}
\end{equation}
where the despurionized form of $\ckco$ is given explicitly by
\begin{equation}
  \co(m) = \ckco(m,0,0;0) \,.
\label{BBOb}
\end{equation}
\end{subequations}
Note that on the left-hand side of Eq.~(\ref{BBOa})
we act with the linearized ST operator $\ckcb$
before despurionization, whereas on the right-hand side we follow
the opposite order.

Since $\ckcb$ and $\cb$ are both linear functional differential operators,
Eq.~(\ref{BBO}) follows using the Leibniz rule
from a corresponding relation that holds at the level of the elementary fields.
When $\f_I(x)$ is one of the (effective) fields
of the target theory, the basic despurionization relation is
\begin{subequations}
\label{BBtilde}
\begin{equation}
  \Big( \ckcb \f_I(x) \Big) \Big|_{despur}
  = \cb \f_I(x) \,,
\label{BBtildea}
\end{equation}
with a similar relation for the $K$ sources of the target theory.
For the spurion-sector fields, the basic relations are
\begin{eqnarray}
  \Big( \ckcb w\f_s(x) \Big) \Big|_{despur}
  &=& \cb m \;=\;  i(2/3)\c m \,.
\label{BBtildeb}\\
  \Big( \ckcb \j_s(x) \Big) \Big|_{despur}
  &=& 0 \,, \qquad\quad
  \Big( \ckcb \bK^{\j_s}(x) \Big) \Big|_{despur}
  \;=\; 0 \,.
\label{BBtildec}
\end{eqnarray}
\end{subequations}
Relations~(\ref{BBtilde}) are established by
exhausting all cases.  As an example,
one has $\ckcb \j_s = \d \ckS_0 / \d \bK^{\j_s}$ and
$\ckcb \bK^{\j_s} = \d \ckS_0 / \d \j_s$, and these expressions,
given in Eq.~(\ref{dS0djs}), were shown to vanish upon despurionization.

With  Eq.~(\ref{BBO}) in hand, Theorem~1(c) follows by using Theorem~1(b)
and noting that when the left-hand side of Eq.~(\ref{BBOa}) vanishes,
the same is true for the right-hand side.  Likewise, Theorem~1(d) follows
directly from Eq.~(\ref{BBO}) and Theorem~1(b).  This completes the proof
of our main result, Theorem~1.

As a final comment we would like to point out the crucial role
of the vanishing right-hand side of Eq.~(\ref{BBtildec}).
The spurion-sector fields $\j_s$ and $\bK^{\j_s}$ leave behind
no remnant in the target theory.  A nonvanishing right-hand side
for Eq.~(\ref{BBtildec}) would leave room for the impossible situation,
mentioned in Sec.~\ref{dyntheory}, of a needed symmetry-restoring counterterm
in the target theory that depends on fields that do not exist in the theory.
As an example, the variation $\ckcb \j_s$ involves
the nonlinear term $w\f_+\f_-$ that depends on the target-theory's
fields only.  But thanks to the factor of $w$, this term disappears
when the $w\to 0$ limit is taken as part of the despurionization process.

\subsection{\label{smr} Summary}
MPW's theorem \cite{MPW1} establishes that
massless supersymmetric gauge theories
are free of anomalies, including in particular supersymmetry anomalies,
provided that the chiral gauge symmetry
of the theory (if it has one) is not anomalous.
We have expanded the scope of the theorem to cover the most general
$\cn=1$ supersymmetric gauge theory,
where the superpotential~(\ref{W}) can contain
parameters with mass dimension one or two.  Each mass parameter $M_{ij}$
was promoted to a dynamical spurion multiplet.
The spurionized theory is under the scope of MPW's original
theorem, and symmetry-restoring counterterms can always
be found order by order.
The generalization to the case that the dimension-two parameters $\O_i$
are present was handled in App.~\ref{Omega} by exhausting all possibilities.

The renormalized target theory is recovered
by despurionization: the new coupling constants that control the coupling
of the (dynamical) spurions to the original fields of the target theory
are sent to zero, while the VEVs of the (scalar) spurions are kept
at values that reproduce the original mass parameters.
With the renormalized action of the spurionized theory in hand,
we showed that despurionization
produces a renormalized action that---as required---depends only on the fields
and parameters of the target theory, and which satisfies all
the ST identities order by order in the quantum target theory.

Referring to the example theory with a single mass parameter
introduced in Sec.~\ref{dyntheory},
when dealing with the effective fields on which the spurionized
1PI functional depends, we have, in particular,
set the scalar spurion field to $m/w$.
In comparison to Eq.~(\ref{expandphis}) that deals with the dynamical field,
this amounts to dropping the quantum part, $\d\f_s$.
The consistency of our treatment of dynamical and effective spurion
fields follows from the physics of the $w\to 0$ limit.
In this limit not only do diagrams with internal spurion lines
vanish; the same is true for all diagrams with
(the quantum part of) the scalar, or the fermion, spurion fields
on an external leg.  In short, for $w\to 0$ the spurion sector
turns into a decoupled, free, massless Wess--Zumino model.
In Sec.~\ref{dyntheory} we already made this observation when discussing the
classical theory; now we have extended it to all orders in perturbation
theory. The only remnant the spurions leave behind is the mass parameter $m$,
which comes from the VEV of $\f_s$.

Our construction has the following implication for
the candidate anomaly of Eq.~(\ref{photinoan}).
The dimensionful coefficients are comprised of two pieces.
We have shown in Sec.~\ref{photino} that the part cubic in the mass parameters,
$a_{ijklmn} M_{ij} M_{kl} M_{mn}$, cannot be altered by any counterterm.
But we have now also proved that a choice of counterterms
always exists such that any
breaking terms in the target theory are eliminated.
Therefore, any concrete diagrammatic calculation must give rise
to a vanishing $a_{ijklmn}$, at every order.
The details regarding the $b_{ijk} M_{ij}\O_k$ part of the
dimensionful coefficient are somewhat different.
Unlike $a_{ijklmn}$, the value of $b_{ijk}$ can be modified by a counterterm
whose structure was discussed in Sec.~\ref{photino}.
Thus, it is not ruled out that a nonzero $b_{ijk}$ is found in a concrete
calculation at some order, but, in that case, $\db^{(n)}$ would necessarily
contain additional terms such that, altogether, $\db^{(n)}$ is $\cb$-exact.

The algebraic mechanism by which the spurionized theory ``deals'' with
the candidate anomaly~(\ref{photinoan}) is investigated in App.~\ref{sspace}.
The spurionized theory's cohomology space contains a nontrivial element,
given explicitly by Eq.~(\ref{Doff}), which reduces to Eq.~(\ref{photinoan})
upon despurionization.  That the candidate anomaly~(\ref{Doff})
will actually never occur in the spurionized theory then follows from its
dependence on the ghost field $c(x)$.
In the case of a vector-like theory, where a gauge invariant
regulator is available, such an anomalous divergence
of the supersymmetry current evidently
cannot arise, because the supersymmetry current is gauge invariant.
In a chiral gauge theory, too, the candidate anomaly~(\ref{Doff}) is ruled out
by invoking the Landau gauge (see Sec.~\ref{mpw1rev}), where the breaking
term can only depend on $\pa\m c(x)$.
The absence of the candidate anomaly~(\ref{photinoan}) from the target
theory is a consequence of the absence of the candidate anomaly~(\ref{Doff})
from the spurionized theory.

The off-shell and on-shell component formalisms differ only in that
additional auxiliary fields are present in the former case.
Their physical contents should be identical.
Nevertheless, in our proof, we had to make use of the on-shell formalism,
because only in that case is an explicit proof available for the massless
case \cite{MPW1}.
In comparison to the spurions used in Ref.~\cite{MPW2,HKS},
an important algebraic difference is that we retrieve a dimensionful
parameter of the original theory from the VEV of the lowest, rather than
the highest, component of the spurion supermultiplet.  A consequence is that
the on-shell supersymmetry transformation rules of our spurion multiplet,
which is just an ordinary chiral supermultiplet, are no longer linear.
As explained in Sec.~\ref{outline}, in order that standard results,
and, in particular, the locality of the breaking terms, will hold
under these circumstances, we have promoted the spurions to new
dynamical fields.\footnote{
  Had a proof been available for the off-shell massless
  theory, our task would have been easier in that
  an off-shell chiral multiplet transforms linearly,
  and this would have allowed us to
  keep the spurions as external fields.
  See, however, Ref.~\cite{JJW} for complications of renormalizing
  supersymmetric theories in the off-shell formalism.
}
As it turns out, the target theory can still be recovered,
essentially because the new Yukawa couplings
we have introduced provide enough control over the coupling between
the dynamical-spurion sector and the physical fields of the original
massive theory.

\section{\label{conc} Conclusion}
In this paper we completed the proof that no anomalies occur
in supersymmetric gauge theories to all orders in perturbation
theory, if their fermion representation is anomaly-free with respect
to the chiral gauge symmetries in the theory.  The proof was given in
the on-shell component formulation, with all auxiliary fields removed.

Such a proof was given before in Ref.~\cite{MPW1} for massless theories,
\ie, theories containing only dimensionless parameters,
also in the on-shell component formalism.
MPW's proof \cite{MPW1} relies on the existence of a generalized BRST
symmetry, which encompasses gauge symmetry, supersymmetry, translation
invariance, and $R$ symmetry.  This makes it possible to turn the subject
into a cohomology problem, which was then solved by MPW for the massless
case, with the result that no anomalies other than the standard chiral
anomaly can occur to all orders in perturbation theory.

The extension of this proof to the most general supersymmetric
gauge theory containing also parameters with positive mass dimension is
not trivial for several reasons.  To begin with, we showed that candidate
anomalies exist that would vanish in the massless case, simply because
they are proportional to dimensionful parameters of the theory.  Technically,
this means that operators with all the right quantum numbers exist that
are closed with respect to the generalized BRST operator, but not exact:
They satisfy the Wess--Zumino consistency conditions,
Eq.~(\ref{closed}), but are not
removable by counterterms.  The task is thus to show that such operators
simply never occur, even if the regulator breaks the symmetries of the
classical theory.

Our strategy was to promote the massive theory to a
theory in which the mass parameters are replaced by spurion multiplets.
The spurionized theory is then a massless theory, to which MPW's theorem
applies.  Usually, spurion fields are kept external;
they appear only on the external legs of diagrams.
In contrast, a key new element of our construction is
that the spurions are made into dynamical fields.
The order-by-order proof of MPW's theorem relies
on locality of the breaking term
that exhibits the violations of Slavnov--Taylor identities.
Locality is a consequence of the Regularized Action Principle,
Eq.~(\ref{limDbreaking}),
which in turn is a manifestation of the quantum equations of motion.
When all fields transforming under the BRST operator are dynamical,
\ie, integration variables in the path integral, the necessary conditions
for MPW's theorem are naturally satisfied.

The theory with
dynamical spurion fields contains diagrams that are not present in the
target theory, because spurion fields can appear on internal lines.
The question is how to ``despurionize,'' in
order to return to the massive target theory.
Using filtration techniques, we showed that this can be done by taking
the couplings of the spurion sector to the physical sector (of the
target theory) to zero.  In the process, the scalar spurions' VEVs
are adjusted so as to reproduce the original mass parameters.

Our analysis establishes that all terms that may break the
generalized BRST symmetry in the quantum theory can, in fact, be removed
by counterterms.  Cohomologically nontrivial operators of the target theory,
which naively could appear as anomalies, will in fact never arise
in any concrete calculation.
Their presence is ruled out by the larger symmetry group
of the spurionized theory,
where the full set of algebraic constraint that are to be satisfied
by the breaking term becomes more powerful.

By construction, we obtain the counterterms in the target theory as descendants
from counterterms in the spurionized theory.
An obvious, but nontrivial, point here is that all the counterterms
needed in the target theory will survive the process of despurionization.
Technically, the violations of Slavnov--Taylor identities are
removable if and only if they can be reproduced by
applying the linearized Slavnov--Taylor operator to
some counterterms.  What makes our construction work
is that the application of the Slavnov--Taylor operator (to the renormalized
1PI functional) and of the linearized Slavnov--Taylor operator (to the
counterterm action or to the breaking term)
both commute with the process of despurionization,
\seef\ Sec.~\ref{despur}.
Therefore, the violations of Slavnov--Taylor identities in the target theory
are completely removed by counterterms that depend only on the fields
and parameters of the target theory itself.

Our construction gives rise to a conserved supersymmetry current
in the target theory, as well as in the spurionized theory,
because the spurions are dynamical fields.
By contrast, the spurion fields of
Refs.~\cite{MPW2,HKS} are external, and, while one can formally use
them to make a breaking term cohomologically exact,
still the supersymmetry current may not be conserved.
As we explain in detail in App.~\ref{MPW2spr}, the basic reason is that
the continuity equation is derived by varying dynamical fields only.

In our proof, we did not consider
soft explicit supersymmetry breaking.  Rather,
starting from a classically supersymmetric theory, we dealt exclusively
with the restoration of supersymmetry along with all other classical symmetries
in the quantum theory.
Once this chapter is accomplished, the usual spurionic techniques
of Refs.~\cite{GG,MPW2,HKS} can then be used in order
to deal with the renormalization of the
soft supersymmetry-breaking terms that are induced in the low-energy
theory as a consequence of the spontaneous breaking of supersymmetry
at a high scale.

With this paper, the proof that there are no perturbative anomalies
in supersymmetric gauge theories other than the usual chiral anomaly
is now complete.

\vspace{3ex}
\noindent {\bf Acknowledgments}
\vspace{3ex}

We thank Aharon Casher and Adam Schwimmer for discussions.
MG thanks the School of Physics and Astronomy of Tel-Aviv University,
and YS thanks the Department of Physics and Astronomy of San Francisco State
University for hospitality.
This work was supported by the Israel Science Foundation under grants
no.~173/05 and 423/09, and, in part, by the US Department of Energy.

\appendix
\section{\label{notation} Notation}
The euclidean Dirac matrices are hermitian.
We use the chiral representation
\begin{equation}
  \g_k =
  \left(
  \begin{array}{cc}
    0      & i\s_k  \\
    -i\s_k & 0
  \end{array}
  \right)
\,,\qquad
  \g_4 =
  \left(
  \begin{array}{cc}
    0  &  1  \\
    1  &  0
  \end{array}
  \right) \,,
\label{4dDirac}
\end{equation}
with $\s_k,$ $k=1,2,3,$ the Pauli matrices.  The chiral projectors are
$P_\pm = (1\pm\g_5)/2$, where
\begin{equation}
  \g_5 = -\g_1\g_2\g_3\g_4
  = \left(
  \begin{array}{cc}
    1  &  0  \\
    0  & -1
  \end{array}
  \right) \,,
\label{pm5}
\end{equation}
and
\begin{equation}
\s_{\m\n}=\frac{i}{2}[\g_\m,\g_\n]\,.
\end{equation}
The charge conjugation matrix $C=-\g_2\g_4$ satisfies
\begin{equation}
  C \g_\m = - \g_\m^T C \,.
\label{C}
\end{equation}

In this paper we use the following Majorana(-like) notation.
Given a two component Weyl fermion $\J$ we construct the
four-component spinors
\begin{equation}
  \j \equiv
  \left(
  \begin{array}{c}
    \vspace{1ex}
    \J \\
    \e\, \bJ^T
  \end{array}
  \right) \,, \qquad
  \bj \equiv
  \left(
  \begin{array}{cc}
    \! -\J^T \e  &  \bJ
  \end{array}
  \right) \,,
\label{maj}
\end{equation}
satisfying the identity
\begin{equation}
  \bj = \j^T \, C  \,.
\label{majcond}
\end{equation}
The following related shorthands are used.
Given a complex scalar field $\f_i$, we let
\begin{eqnarray}
  \f_{5i}
  & \equiv &
  P_+ \f_i + P_- \f^*_i
  = \left(
  \begin{array}{cc}
    \f_i  &  0  \\
    0   &  \f_i^*
  \end{array}
  \right) \,.
\nonumber
\end{eqnarray}
For the group generators and covariant derivatives we similarly set
\begin{eqnarray}
  T_{5a}
  & \equiv & \rule{0mm}{8mm}
  \left(
  \begin{array}{cc}
    T_a  &  0  \\
    0   &  -T_a^*
  \end{array}
  \right)
  =
  \left(
  \begin{array}{cc}
    T_a  &  0  \\
    0   &  -T_a^T
  \end{array}
  \right) \,,
\NON
  D_{5\m}
  & \equiv & \rule{0mm}{8mm}
  (\partial_\m +igA_{\m a} T_{5a})
  =
  \left(
  \begin{array}{cc}
    D_\m  &  0  \\
    0   &  D_\m^*
  \end{array}
  \right) \,.
\nonumber
\end{eqnarray}
For example, $D_{5\m} \f_5$ sandwiched between two spinors is $D_\m\f$ or
$D_\m^*\f^*$, depending on the chirality of the spinors.
If $W=W(\f_i)$ is a polynomial in a set of complex scalar fields $\f_i$,
we similarly define $W_5 = P_+ W + P_- W^*$, as well as
$W_{5,i} = P_+\, \partial W / \partial \f_i
+ P_-\, \partial W^* / \partial \f_i^*$, and so on.
Because of the presence of $\g_5$, the Dirac matrices
don't commute with $D_{5\m}$, and, in particular,
$\g_\m D_{5\m} = D^*_{5\m} \g_\m $.  We will also use the shorthands
\begin{equation}
  \Sl{D}_5  \equiv \g_\m D_{5\m}  \,,
\qquad
  \Sl{D}^*_5  \equiv \g_\m D^*_{5\m} \,,
\label{D5}
\end{equation}
in which the Dirac matrix always occurs to the left of the derivative
operator.

\section{\label{offshell} Off-shell formalism}
In this appendix we discuss the BRST operator for the off-shell
component formalism.  We elaborate on the connection with the on-shell
classical action, including in particular its $K$ source terms,
and show that the on-shell action satisfies the ST identity.

In the off-shell framework, the gauge multiplet
consists of $(A_{\m a},\l_a,D_a)$ and the matter multiplet of  $(\f_i,\j_i,F_i)$,
where $D_a$ and $F_i$ are auxiliary fields.
The off-shell lagrangian consists of separate kinetic terms
for the gauge multiplet\footnote{
  See App.~\ref{notation} for notation.
}
\begin{equation}
  \cl_g = {1\over 4} F_{\m\n a}^2 + \half \bl_a \Sl{D}_{ab} \l_b + D_a^2 \,,
\label{Lgauge}
\end{equation}
and for each matter multiplet,
\begin{equation}
  \cl_k = (D_\m \f)^*_j (D_\m \f)_j + ig D_a \f^*_i T_{aij} \f_j + F^*_i F_i
  + {1\over 2} \bj_i \Sl{D}_{5ij} \j_j
  - ig\sqrt{2}\, \bl_a \f^*_{5i} T_{5aij} \j_j \,,
\label{Lk}
\end{equation}
as well as a superpotential-dependent term
\begin{equation}
  \cl_p = -i\, W_{,i}\, F_i
          -i\, W_{,i}^*\, F^*_i
  + {1\over 2}\, \bj_i\, W_{5,i,j} \, \j_j \,.
\label{Lp}
\end{equation}
The gauge-fixing action~(\ref{Segf}) remains the same as in the on-shell case.

The supersymmetry transformations are linear in superspace.\footnote{
  For a discussion of the superspace cohomology see App.~\ref{sspace}.
}
In the off-shell component formalism they become nonlinear:
the original, linear supersymmetry transformation
is accompanied by a supergauge transformation.  The latter
restores the Wess--Zumino gauge, where unphysical components of
the vector superfields are eliminated algebraically.
This entails several changes:
Ordinary derivatives turn into covariant ones, and
the transformation rule of any nonsinglet $F_i$ picks up an additional,
nonlinear term that involves the gaugino.  Explicitly,
the off-shell transformation rules of the gauge multiplet are
\begin{eqnarray}
  sA_{\m a} &=& \bh\g_\m\l_a + D_{\m ab}c_b + \x_\n\pa{\n}A_{\m a}
\,,\label{gaugeBRST}
\\
  s\l_a &=& \Big( (i/2) F_{\m\n a}\s_{\m\n} + D_a\g_5 \Big)\h
  + g f_{abc} c_b \l_c + \x_\n\pa{\n}\l_a + i\c \g_5 \l_a \,,
\nonumber\\
  sD_a &=& \rule{0ex}{3ex}
         -\bh\g_5\Sl{D}_{ab} \l_b + g f_{abc} c_b D_c + \x_\n\pa{\n}D_a \,.
\nonumber
\end{eqnarray}
For the matter multiplets they are
\begin{eqnarray}
  s\f_i &=& \sqrt{2}\,\bh P_+\j_i - ig c_a T_{aij}\f_j + \x_\n\pa{\n}\f_i
  +i (2/3) \c \f_i \,,
\label{matterBRST}
\\
  s\f_i^* &=& \sqrt{2}\,\bh P_-\j_i + ig c_a T_{aij}^*\f_j^*
  + \x_\n\pa{\n}\f_i^* - i (2/3) \c \f_i^* \,,
\nonumber\\
  s\j_i & = & \sqrt{2}\left(\Sl{D}_{5ij}^*\f_{5j}^* + iF_{5i} \right)\h
  - ig c_a T_{5aij}\j_j + \x_\n\pa{\n}\j_i -i(1/3)\c \g_5\j_i \,,
\nonumber\\
  sF_i &=& -i\sqrt{2}\,\bh P_-\Sl{D}_{ij} \j_j
           + 2g\,\bh P_-\l_a T_{aij}\,\f_j
           -igc_a T_{aij} F_j + \x_\n\pa{\n}F_i -i(4/3)\c F_i\,,
\nonumber\\
  sF^*_i &=& -i\sqrt{2}\,\bh P_+\Sl{D}_{ij}^* \j_j
           - 2g\,\bh P_+\l_a T_{aij}^*\,\f^*_j
           + igc_a T_{aij}^*F^*_j + \x_\n\pa{\n}F^*_i +i(4/3)\c F^*_i\,.
\nonumber
\end{eqnarray}
Note that the off-shell transformation rules of the boson fields
$A_\m$ and $\f_i$ remain the same as in  Sec.~\ref{onsh}.
The transformation rules of the ghost-sector fields and of parameters
are unchanged, too.

The off-shell BRST operator is nilpotent, $s^2=0$.  Splitting up the off-shell
operator similarly to Eq.~(\ref{ssplit}) we have
\begin{equation}
  0 = s_g^2 = s_\x^2 = s_\c^2 = \{s_g,s_\x\} = \{s_g,s_\c\} = \{s_\x,s_\c\}
   = \{s_\h,s_\c\} \,.
\label{salg}
\end{equation}
In addition we have $(s_g + s_\x + s_\h)^2=0$, which, together with
the above relations, may be rewritten as
\begin{equation}
  s_\h^2 = -\{s_\h,s_g\} - \{s_\h,s_\x\}\,,
\label{ssalg}
\end{equation}
showing that two off-shell supersymmetry transformations
close on the sum of a translation and
a gauge transformation.  The local parameter of this gauge transformation
is $\propto \bh \Sl{A}\h$,
making the sum of the two terms on the right-hand side
a covariant translation with parameter $\bh \g_\m \h$.

As mentioned in Sec.~\ref{ARrev}, in the off-shell formalism
the $K$-source action takes the form of Eq.~(\ref{sext}), and no terms
bilinear in the $K$ sources exist.  That the off-shell classical action
satisfies the ST identity follows from nilpotency of
the off-shell transformation.

We next discuss the transition from the off-shell to the on-shell
formalism.  In the rest of this appendix, we use superscripts to distinguish
objects pertaining to the on- or off-shell formalisms.
The transition is facilitated by
integrating out the auxiliary fields $F_i$, $F_i^*$ and $D_a$,
while turning off their sources $K^F_i$, $K^{F^*}_i$ and $K^D_a$.
The full on-shell classical action is thus obtained as
\begin{equation}
  \exp(-\Scl^{on}) = \int \prod_a \cd D_a \prod_i \cd F_i \cd F_i^*\;
  \exp(-\Scl^{off}) \bigg|_{K^F_i=K^{F^*}_i=K^D_a=0} \,.
\label{Sonshell}
\end{equation}
Integrating out the auxiliary fields has several effects.
The auxiliary fields occurring in the off-shell supersymmetric action
and in the supersymmetry transformation
rules of the fermions are replaced by the expressions they are
equal to by their equations of motion.
In particular, terms bilinear in the fermionic $K$-sources are generated.
Their explicit form is
\begin{equation}
  S_{ext}^{on,bil} =  2 (\bK_i^{\j} P_+ \h)(\bK_i^{\j} P_- \h)
  -\frac{1}{2} (\bK_a^{\l} \g_5 \h)^2 \,.
\label{bilK}
\end{equation}
This leads to a difference between the on-shell $\cb^{on}$ and $s^{on}$,
when acting on a fermion field:
\begin{subequations}
\label{Bs}
\begin{eqnarray}
  (\cb^{on}-s^{on}) \l_a &=& - \g_5 \h\, (\bK_a^{\l} \g_5 \h) \,,
\label{Bsa}\\
  (\cb^{on}-s^{on}) \j_i &=& 2 P_+ \h\, (\bK_i^{\j} P_- \h)
   +2 P_- \h\, (\bK_i^{\j} P_+ \h)\,.
\label{Bsb}
\end{eqnarray}
\end{subequations}
Using its definition~(\ref{Sonshell}) one can show
that $\Scl^{on}$ satisfies the ST identity, $\cs(\Scl^{on})=0$.
First, consider the term in
the on-shell ST identity that corresponds to a scalar field $\f_i$.
With expectation values that refer to the auxiliary-field partition
function~(\ref{Sonshell}) we have
\begin{subequations}
\label{offon}
\begin{eqnarray}
  \frac{\d \Scl^{on}}{\d \f_i} \frac{\d \Scl^{on}}{\d K^\f_i}
  &=&
  \svev{\frac{\d \Scl^{off}}{\d \f_i}} \svev{\frac{\d \Scl^{off}}{\d K^\f_i}}
\label{offona}\\
  &=&
  \svev{\frac{\d \Scl^{off}}{\d \f_i}\frac{\d \Scl^{off}}{\d K^\f_i}}
\label{offonb}\\
  &=& \svev{s^{off} \f_i \; \frac{\d \Scl^{off}}{\d \f_i}} \,.
\label{offonc}
\end{eqnarray}
\end{subequations}
The transition from the first to the second line works as follows.
On the first line, because $\d \Scl^{off} / \d K^\f_i$ is independent of
the auxiliary fields, we can drop the expectation value surrounding it:
$\svev{\d \Scl^{off} / \d K^\f_i} = \d \Scl^{off} / \d K^\f_i$.
For the same reason, $\d \Scl^{off} / \d K^\f_i$ can now be brought inside
the the expectation value $\svev{\d \Scl^{off} / \d \f_i}$, obtaining
Eq.~(\ref{offonb}).
The last equality follows by noting that
$\d \Scl^{off} / \d K^\f_i = s^{off}  \f_i$.
A similar equality holds for every dynamical field present in the on-shell
formalism.  Which term can be
moved inside and outside of the expectation values varies,
but the outcome is the same.
In addition, one can check that
$\svev{s^{off} F_i \, (\d \Scl^{off} / \d F_i)}=0$,
with a similar result for $D_a$, which is true because
the auxiliary-field equations of motion can be used inside the
averages of Eq.~(\ref{Sonshell}).
Finally, by summing $\svev{s^{off} \F_I \, (\d \Scl^{off} / \d \F_I)}$
over all of the dynamical fields, now those of the off-shell formulation,
we find that $\cs(\Scl^{on})= \svev{s^{off} \Scl^{off}} = 0$, as claimed.

The construction can be generalized to relate off-shell and on-shell
cohomologically closed solutions. This is based on the following trick.
Assume that the commuting functional $\G$ satisfies the ST identity,
$\cs(\G)=0$, and that the anti-commuting functional $\D$
is closed relative to the corresponding linearized ST operator,
namely, $\cs_\G \D = 0$.
Introducing a grassmann variable $\z$ we then have
\begin{equation}
  \cs(\G+\z\D) = \cs(\G) + \z\,\cs_\G \D = 0 \,,
\label{Gt}
\end{equation}
where we used that $\z^2=0$.
We now assume that $\cb^{off} \db^{off} = 0$, and seek
the corresponding on-shell closed solution that
satisfies $\cb^{on} \db^{on} = 0$.
(Recall $\cb^{off} = \cs_{\Scl^{off}}$ and $\cb^{on} = \cs_{\Scl^{on}}$.)
Augmenting the auxiliary-field partition function~(\ref{Sonshell})
by the grassmann variable $\z$, we define
\begin{eqnarray}
  Z(\z) &\equiv&  \prod_a \cd D_a \prod_i \cd F_i \cd F_i^*\;
  \exp(-\Scl^{off}-\z\db^{off})
\label{Donshell}\\
  &=& \exp(-\Scl^{on}-\z\db^{on}) \,,
\nonumber
\end{eqnarray}
which implies
\begin{equation}
  \db^{on} = -\frac{\partial}{\partial \z} \log Z(\z)
  = \svev{\db^{off}} \,,
\label{dondef}
\end{equation}
where $\z^2=0$ was used.
The expectation value in Eq.~(\ref{dondef}) is with respect to the
partition function~(\ref{Sonshell}), \ie, with respect to $Z(\z=0)$.
Equation~\ref{dondef} has the intuitively expected structure, namely,
the on-shell solution is obtained from the off-shell one
by substituting for the auxiliary fields using their equations of motion
(in the presence of $K$ sources for the (bosons and) fermions).

In order to show that the definition~(\ref{dondef})
satisfies $\cb^{on} \db^{on} = 0$ one uses Eq.~(\ref{Gt})
as well as equalities similar to Eq.~(\ref{offon}) in which
$\Scl$ is replaced with $\Scl+\z \db$, and expectation values
are now with respect to $Z(\z)$.
However, a word of caution is that the necessary intermediate steps
depend on the detailed form of $\db^{off}$, and must be verified on a
case by case basis.  This refers to the ability to maneuver
at least one of $\d (\Scl^{off} + \z\db^{off}) / \d \F_I$
or $\d (\Scl^{off} + \z\db^{off}) / \d K_I$
inside and outside of the expectation values.

\section{\label{Omega} Ruling out $\O$-dependent anomalies}
A soft anomaly that is linear in the dimension-two parameters $\O_k$
will have mass dimension three, when it is independent
of all mass parameters, or mass dimension two, when in addition it is linear
in the mass parameters $M_{ij}$.\footnote{
  No operators with mass dimension one and the right quantum numbers exists.
}
Most of this appendix is devoted to excluding
the dimension-three case.  But first let us consider the dimension-two case.

The dimension-two operator must have the form of $\h$ times
a fermion field.  Both $\bh P_\pm \j_i$,
with $\j_i$ a neutral matter fermion, and $\bh P_\pm \l$,
with $\l$ an abelian gaugino, are cohomologically closed.\footnote{
  Provided that they are multiplied by dimensionful coefficients
  with an appropriate $R$ charge, as in Eq.~(\ref{photinoan}).
}
The first is exact, being the supersymmetry transformation of the corresponding
scalar, $\f_i$ or $\f_i^*$.

The operators $\bh P_\pm \l$ are not exact,
but they are ruled out by generalizing
the construction of Sec.~\ref{spur}.  We spurionize the mass parameters,
then renormalize the spurionized theory such that its breaking term
vanishes to all orders, and finally, via despurionization,
we reconstruct the quantum target theory while preserving
all of its classical symmetries.  This procedure will be successful
provided that we can handle the case where
the only dimensionful parameters present are the $\O_k$'s.

With no mass parameters $M_{ij}$ in the superpotential, the most general form
of the breaking term is
\begin{equation}
  \db' = \tdb + (\O_k X'_k +\hc) \,,
\label{Okprime}
\end{equation}
where the $\O_k$ dependence is explicitly shown,
and summation over $k$ is implied.   Turning off momentarily
all the $\O$'s, it follows from MPW's theorem that there exist a counterterm $Q$
(which in itself is independent of $\O_k$) such that $\tdb = -\cb|_{\O_k=0}\, Q$.
Turning the $\O$'s back on, it follows that
$\tdb + \cb Q = (\cb - \cb|_{\O_k=0})Q$ is linear in $\O_k$.
The breaking term thus has another representative $\db=\db'+\cb Q$
such that
\begin{equation}
  \db = \O_k X_k +\hc = \int d^4x\, \O_k X_k(x) +\hc \,,
\label{Ok}
\end{equation}
which now contains no $\O$-independent terms.

In the rest of this appendix we show that there exists no
cohomologically nontrivial solution with the form of Eq.~(\ref{Ok})
that also satisfies all the other required constraints.
We begin by listing these constraints.

In order that $\db$ would qualify as a candidate breaking term,
$X_k(x)$ must have mass dimension
$\cq_d=3$, $R$ charge $\cq_R = -4/3$, and ghost number $\cqgh = 1$.
The latter requirement implies that $X_k(x)$ should be (at least)
linear in $\h$ or $c(x)$.  However, as reviewed in Sec.~\ref{mpw1rev},
any $c(x)$ dependence can be only through $\partial_\m c(x)$.
Therefore we must have $X_k(x) = Y_{k\m}(x) \partial_\m c(x)$,
where $Y_{k\m}(x)$ has $\cq_d = 1$, and $\cq_R = -4/3$.
Such an object does not exist, ruling out anything depending on $c(x)$.

Below we will write $X_k$, suppressing
the dependence on $x$.  Whether we refer to the local operator
or to the integrated one will be clear from the context.

Operators with more than one object with nonzero ghost number
(for a complete list, see Sec.~\ref{onsh}) are ruled out.
As an example, we consider the case of terms with two $\h$'s.
A bilinear in $\h$ must contain $\g_\m$ or $\s_{\m\n}$.
In the latter case, the only $\cq_d=3$ operator is
$(\bh \s_{\m\n} \h) F_{\m\n}$, which
has the wrong ghost number and $R$ charge.  For the other case,
we can have  $(\bh \g_\m \h) K_\m^A$, where $K_\m^A$
is the source coupled to the variation of an abelian gauge field.
Now the ghost number is correct, but the $R$ charge is wrong.

Having concluded that $X_k$ must be linear in $\h$ we may write
$X_k = \bh (P_+ Y_{k+} + P_- Y_{k-})$.
Then $\cq_d(Y_{k\pm}) = 5/2$ while the $R$ charge is $\cq_R(Y_{k+}) = -7/3$
and $\cq_R(Y_{k-}) = -1/3$.  Either way, $Y_{k\pm}$ must contain a field
with half-integer dimension, \ie, a matter fermion $\j$, a gaugino $\l$,
or the $K$ source for one of them.  This must be multiplied by an object
with $\cq_d=1$, which can be the gauge field, a scalar field,
or a derivative.
These possibilities do not sum up to anything with $\cq_R = -7/3$,
thereby ruling out $Y_{k+}$.  Dropping the minus subscript from now on,
it follows that $X_k = \bh P_- Y_k$ for some $Y_k$.

Taking into account all possibilities for the operator $Y_k$,
we arrive at the most general expression
allowed by dimensionality, ghost number, \etc:
\begin{equation}
  X_k = c_{1,ik}\, \bh P_- \Sl{A}_{ij} \j_j
  + c_{2,ijk}\,  \f^*_i (\bh P_- \j_j) + c_{3,iak}\, \f_i (\bh P_- \l_a) \,,
\label{allX}
\end{equation}
where the indices $i,j$ run over the matter supermultiplets.
We have dropped terms $\propto \bh P_- \sl\partial \j_i$
since they are total derivatives.

Because $X_k$ is independent of the ghost-sector fields, $\cb$-closedness
implies that $\cb_g X_k$ vanishes separately for each $k$,
and so does $\cb_\h (\O_k X_k)$ after the $k$-summation.
The breakup of $\cb$ is defined analogously to Eq.~(\ref{ssplit}).  In particular,
$\cb_\h$ consists of the $\h$-dependent terms in the variation
of each field.  For the operators that occur in Eq.~(\ref{allX}), $\cb_g = s_g$.
Note that $\cb_\h \ne s_\h$ when acting on a fermion field.

We start from the requirement that $\cb_g X_k = s_g X_k = 0$.  Once again,
since $X$ is independent of the ghost-sector fields, $s_g X_k = 0$ implies
that $X_k$ is gauge invariant.\footnote{
  Recall that, by gauge invariance of the superpotential~(\ref{W}),
  $\O_k$ can be nonzero only if $\f_k$ is a gauge singlet.
}
  This rules out the first term in Eq.~(\ref{allX}).
We comment that, by adding a total-derivative term, one can replace
$A_\m$ by a covariant derivative $D_\m$ in that term.  While the resulting
operator would transform covariantly, it will never be gauge invariant.

For the last two terms, we find that $c_{2,ijk}$ can be nonzero
only when $i,j$ correspond to two complex-conjugate representations.
As for $c_{3,iak}$, it can be nonzero only if $\f_i$ belongs
to the adjoint representation in the nonabelian case,
or is neutral in the abelian case.

We next turn to $\cb_\h$.  The variation of the last term
in Eq.~(\ref{allX}) gives rise  to
$c_{3,iak}\, (\bh P_- \l_a)(\bh P_+ \j_i)$, among other terms.
This does not vanish by fierzing,
and cannot cancel against any other term in $\cb_\h\, \db$,
ruling out this possibility.

It remains to consider the $c_2$ term.   Explicitly, we have
\begin{eqnarray}
  \cb_\h\, c_{2,ijk}\, \f^*_i (\bh P_- \j_j)
  &=&
  \sqrt{2} c_{2,ijk} \left( (\bh P_-  \j_i)(\bh P_-  \j_j)
  + \f_i^* (\bh P_- (\Sl{D} \f)_j^* \h) \right)
\NON
  &=&
  \sqrt{2} c_{2,ijk} \left[ (\bh P_-  \j_i)(\bh P_-  \j_j)
  + \frac{1}{4} (\bh \g_\m\h) \partial_\m (\f_i^*  \f_j^*) \right.
\NON
  && \left. + \frac{1}{4}
     (\bh \g_\m\h) (\f_i (D_\m \f_j) - \f_j (D_\m \f_i))^* \right] \,.
\label{varc3}
\end{eqnarray}
Disregarding the total derivative, the expression in square brackets
on the last line is anti-symmetric in the indices $i$ and $j$.
The required vanishing of the left-hand side thus imposes the
constraint that $c_{2,ijk}$ is symmetric in the indices $i,j$.
This, in turn, implies
\begin{equation}
  c_{2,ijk}\, \O_k \f^*_i (\bh P_- \j_j)
  = 2^{-3/2} c_{2,ijk}\, \O_k \cb_\h (\f^*_i \f^*_j)
  = 2^{-3/2} c_{2,ijk}\, \cb (\O_k \f^*_i \f^*_j) \,,
\label{Xexact}
\end{equation}
hence the only allowed term is $\cb$-exact.

\section{\label{embed} Embedding lemma}
Here we give the proof of Lemma~2.
We start by applying the filtration $\cn$ to the closedness
relation $BX =0$.  From the lowest-$\cn$ term it follows that
$B_{\und{0}} X_{\und{k}} =0$,
\ie, $X_{\und{k}}$ belongs to the cohomology of $B_{\und{0}}$
with the same quantum numbers.  If $X_{\und{k}}$ is $B_{\und{0}}$ nontrivial,
we are done.  If not, there exists $Q_{\und{k}}$ such that
$B_{\und{0}} Q_{\und{k}} = X_{\und{k}}$. It follows that
$X^{(1)} = X - B Q_{\und{k}}$ is a representative of the same $B$ equivalence
class and that, moreover, $X^{(1)}$
has a lowest-$\cn$ value equal to (or greater than) $k+1$.
Note that $X^{(1)}$ may have a largest $\cn$ higher than that of $X$, but under
the assumptions of Lemma~2 we are assured that the highest-$\cn$ value
is always bounded by $n_{max}$.

Now the process is repeated.  Again,
the new lowest-$\cn$ part, $X^{(1)}_{\und{k+1}}$, is $B_{\und{0}}$-closed.\footnote{
  If $X^{(1)}_{\und{k+1}}=0$ then, trivially, $Q_{\und{k+1}}=0$
  and we may proceed to $X^{(1)}_{\und{k+2}}$.
}
If it is also $B_{\und{0}}$-nontrivial then we are done.
If not, there exists $Q_{\und{k+1}}$ such that
$X^{(1)}_{\und{k+1}} = B_{\und{0}} Q_{\und{k+1}}$
and we may repeat the process once more.
Since, at each step, the lowest-$\cn$ part of the representative is raised
by one, and since the maximal value we may encounter is $n_{max}$,
the process must stop.  If it did not stop earlier, we would have constructed
$Q=Q_{\und{k}}+Q_{\und{k+1}}+\cdots+Q_{\und{n_{max}}}$ such that $X - B Q = 0$.
But this would imply that $X$ was $B$-exact,
contrary to our assumption that it is nontrivial.
Therefore the process had to stop earlier,
when the lowest-$\cn$ part
of $X - B (Q_{\und{k}} + Q_{\und{k+1}} + \cdots + Q_{\und{m}})$
is $B_{\und{0}}$-nontrivial (and, necessarily, $m<n_{max}$).

\section{\label{wnct} Avoiding $\cnw<0$ symmetry-restoring counterterms}
In this appendix we prove the following result.
Consider the breaking term $\ckdb$ obtained in the spurionized theory
at some order in the loop expansion.
We assume that $\ckdb$ contains no part with $\cnw<0$, \ie, with
negative $w$-number.\footnote{
  That this assumption is true order by
  order is proved in Sec.~\ref{despur}.
}
Then, a symmetry restoring counterterm $\ckQ$ can be chosen
such that $\ckdb = \ckcb \ckQ$, and $\ckQ$ has no $\cnw<0$ part.

\vskip 1ex

We start with the $w$-number filtration of the nilpotency relation,
$\ckcb^2=0$.  Using Eq.~(\ref{spurwnb}) we obtain
\begin{subequations}
\label{nilBwn}
\begin{eqnarray}
  \cb_{\und{0}}^2 &=& 0 \,,
\label{nilBwna}\\
  \{ \cb_{\und{0}}, \cb_{\und{2}} \}  &=& 0 \,,
\label{nilBwnb}\\
  \cb_{\und{2}}^2 &=& 0 \,.
\label{nilBwnc}
\end{eqnarray}
\end{subequations}
The $w$-number filtration of the breaking term is
$\ckdb = \db_{\und{0}} + \db_{\und{2}} + \cdots + \db_{\und{2n}}$.
Since even and odd values of $\cnw$ don't mix under the action of
$\ckcb$, we have assumed
that $\ckdb$ contains only even powers.  (For odd powers only,
the proof would go the same.)  Our assumptions rule out
that $\db$ has any $\cnw<0$ part.  Also, allowing for the possibility
that individual terms in the expansion vanish,
we may assume without loss of generality that $2n=2n_{max}$ is the maximal
(even) value of $\cnw$ that is allowed by the quantum numbers $\cq_d=1$,
$\cqgh=1$ and the given loop number $\cql$.

We now proceed to the construction of the symmetry-restoring counterterm.
Filtering $\ckcb \ckdb = 0$, we have
\begin{subequations}
\label{Deltawn}
\begin{eqnarray}
  \cb_{\und{0}} \db_{\und{0}} &=& 0 \,,
\label{Deltawna}\\
  \cb_{\und{0}} \db_{\und{2}} +\cb_{\und{2}} \db_{\und{0}} &=& 0 \,,
\label{Deltawnb}\\
  \cb_{\und{0}} \db_{\und{4}} +\cb_{\und{2}} \db_{\und{2}} &=& 0 \,,
\label{Deltawnc}\\
  \vdots &=& \vdots
\NON
  \cb_{\und{0}} \db_{\und{2n_{max}}} +\cb_{\und{2}} \db_{\und{2n_{max}-2}}
  &=& 0 \,,
\label{Deltawnd}\\
  \cb_{\und{2}} \db_{\und{2n_{max}}} &=& 0 \,.
\label{Deltawne}
\end{eqnarray}
\end{subequations}
By Lemma~3, the breaking-term cohomology
of $\cb_{\und{0}}$ is trivial. It follows from Eq.~(\ref{Deltawna}) that
there is $Q_{\und{0}}$ such that
\begin{equation}
  \db_{\und{0}} = \cb_{\und{0}} Q_{\und{0}}\,.
\label{Deltawn0}
\end{equation}
Next, plugging it into Eq.~(\ref{Deltawnb}) we have
\begin{equation}
  0 = \cb_{\und{0}} \db_{\und{2}} +\cb_{\und{2}} \db_{\und{0}}
    = \cb_{\und{0}} \db_{\und{2}} +\cb_{\und{2}} \cb_{\und{0}} Q_{\und{0}}
    = \cb_{\und{0}} (\db_{\und{2}} - \cb_{\und{2}} Q_{\und{0}}) \,,
\label{Deltawn02}
\end{equation}
where in the last step we used Eq.~(\ref{nilBwnb}).
Again using the triviality of the $\cb_{\und{0}}$ cohomology,
there exists  $Q_{\und{2}}$ such that
\begin{subequations}
\label{Deltawn2}
\begin{equation}
   \cb_{\und{0}} Q_{\und{2}} = \db_{\und{2}} - \cb_{\und{2}} Q_{\und{0}} \,,
\label{Deltawn2a}
\end{equation}
or equivalently,
\begin{equation}
  \db_{\und{2}} = \cb_{\und{2}} Q_{\und{0}} + \cb_{\und{0}} Q_{\und{2}} \,.
\label{Deltawn2b}
\end{equation}
\end{subequations}
The next step is to plug Eq.~(\ref{Deltawn2b}) into Eq.~(\ref{Deltawnc}).
We find that $\db_{\und{4}} - \cb_{\und{2}} Q_{\und{2}}$ is closed,
and thus, by Lemma 3,
equal to $\cb_{\und{0}} Q_{\und{4}}$ for some $Q_{\und{4}}$.
This time we have made use of Eq.~(\ref{nilBwnc}) to obtain
$\cb_{\und{2}}^2 \db_{\und{0}} = 0$.

This goes on iteratively, until we reach
\begin{equation}
  \cb_{\und{0}} Q_{\und{2n_{max}}}
  = \db_{\und{2n_{max}}} - \cb_{\und{2}} Q_{\und{2n_{max}-2}} \,.
\label{Deltawnn}
\end{equation}
At this point we have made use of all of
the relations~(\ref{Deltawn}) except the last one, Eq.~(\ref{Deltawne}).
Letting $\ckQ = Q_{\und{0}} + Q_{\und{2}} + \cdots + Q_{\und{2n_{max}}}$
we now have
\begin{equation}
  \ckcb \ckQ =  \ckdb + \cb_{\und{2}} Q_{\und{2n_{max}}} \,.
\label{Deltawnplustwo}
\end{equation}
But $\cb_{\und{2}} Q_{\und{2n_{max}}}$ must in fact vanish.  The reason
is that it has $\cnw=2n_{max}+2$, whereas by assumption the highest possible
$\cn$ allowed for the quantum numbers of $\ckdb$ is $2n_{max}$.  Hence,
\begin{equation}
  \ckcb \ckQ =  \ckdb \,,
\label{DeltawnQ}
\end{equation}
and we have succeeded in constructing the desired counterterm.

\section{\label{sspace} Superspace origin of the abelian gaugino cohomology class}
In this appendix we elaborate on the algebraic origin of the abelian gaugino
(candidate) anomaly found in Sec.~\ref{photino}.  We show that, in superspace,
it can be traced back to a class of abelian supergauge anomalies.
We then explore what shape these cohomologically nontrivial solutions
take in the off-shell component formalism.  This clarifies the crucial role
of the constraint on $c(x)$ dependence, reviewed in Sec.~\ref{mpw1rev},
in excluding the abelian gaugino anomaly via spurion methods.

The advantage of superspace is that the supersymmetry transformations
are linear, and close on an ordinary translation.
The price is that local gauge transformations must be promoted to a larger
local symmetry, parametrized by a chiral superfield.  Consequently,
the superspace formulation contains many new unphysical degrees of freedom.

The BRST setup is likewise simpler in superspace.  Following the notation
of Sec.~\ref{onshell} we denote by $s_\h$ and $s_\x$ the (anti-commuting)
operators that effect supersymmetry transformations and translations
respectively.  The operator $s_\h+s_\x$ is nilpotent.\footnote{
  As in Sec.~\ref{onshell} one has $s_\h\,\x_\m = -\bh\g_\m\h$,
  \seef\ Eq.~(\ref{parona}).
}

Supergauge transformations are effected by another nilpotent BRST operator,
denoted $s_{SG}$ \cite{PSGI}.  Its action on any superfield that occurs in
the invariant superspace lagrangian is obtained from the ordinary supergauge
transformation rules by simply replacing the usual gauge-transformation
superfield by the ghost superfield, hereby denoted $\L$.
For what concerns us, we only need to know
that the action of $s_{SG}$ on a chiral (matter) superfield has the same
form as an ordinary infinitesimal gauge transformation, namely,
$s_{SG}\, \F_i = -ig \L_a T_{aij} \F_j$ (compare the second term on the
right-hand side of Eq.~(\ref{sonc})).  Also, in the abelian case,
the ghost superfield itself is BRST invariant, $s_{SG}\, \L = 0$.

Based on this information we can immediately write down a family
of cohomologically nontrivial solutions, each constituting a (candidate)
supergauge anomaly, given by
\begin{equation}
  \db = \int d^4x\,d^2\th\, \L\F  \,.
\label{sAp}
\end{equation}
Here $\L$ is the abelian ghost chiral superfield and $\F$
is any (super)gauge invariant chiral superfield with zero ghost number.
The ghost number of $\db$, inherited from $\L$, is one,
as it should.

Let us verify that $\db$ of Eq.~(\ref{sAp})
is cohomologically nontrivial.\footnote{
  In this appendix we disregard the $R$ symmetry part of the BRST operator.
}
Since $\db$ involves a (chiral) superspace integral,
it is supersymmetry invariant.  Also, from the BRST supergauge transformation
rules discussed above and the gauge invariance of $\F$
it follows that $s_{SG}\, \db = 0$ as well.
This shows that $\db$ is cohomologically closed.  That it is not
exact basically follows from the fact that $\F$ is gauge invariant,
and thus the action of $s_{SG}$ on $\F$ gives zero, and not $\L\F$ .

There are various possibilities for the chiral superfield $\F$.
It can be a composite superfield which is the (gauge invariant) product
of two or three matter superfields.  Another possibility is that
$\F$ is an elementary, gauge-singlet matter superfield.

We may also set $\F$ to a ($\th$-independent) constant.
In that case, $\db$ collapses to the $F$-component of the ghost superfield,
denoted $\L_{\th\th}$.
Moreover, the cohomology class~(\ref{sAp}) with $\F=1$ has another
representative which is nothing but $\bh P_- \l$ (see Eq.~(\ref{photinoan})).
It is obtained as $\db +(s_\h + s_{SG}) Q$, where the counterterm $Q$ is
related to the scalar and pseudo-scalar $\th\th$ components
of the (abelian) vector superfield $V$.
In the notation of Ref.~\cite{SW},\footnote{
  Where the relevant components of the vector superfield read
  $V = \cdots -(i/2) (\bth \g_5 \th) M - (1/2) (\bth \th) N + \cdots\,,$
  see Chapter 26 therein.
}
we have $Q\propto M-iN$.  Since $s_{SG}(M-iN) \propto \L_{\th\th}$,
and $s_\h(M-iN) \propto \bh P_- \l$, we may choose the coefficient of $Q$
such that the $\L_{\th\th}$ term vanishes, and
the new representative of $\db$ becomes a purely supersymmetry anomaly.

To our knowledge, the most serious attempt to construct a consistent
regularization method in superspace was carried out in Ref.~\cite{bjW}.
Its starting point is the higher-derivative regularization of the Wess--Zumino
model \cite{IZ}.  However, the gauge theory case is significantly more
complicated.  Apart from higher (covariant) derivatives, one must introduce
a set of Pauli--Villars fields.  Furthermore, the regularization proposed
in Ref.~\cite{bjW} is actually a two-cutoff method, where the fully regularized
theory (the ``pre-regulator'' level) does not preserve gauge invariance.
Therefore, a supergauge anomaly is in fact a logical possibility in this
context.

We next discuss how the cohomology classes of Eq.~(\ref{sAp}) are
realized within the off-shell component formalism of App.~\ref{offshell}.
Expanding the ghost superfield as $\L=(\f_\L,\j_\L,F_\L)$,
we replace the components
\begin{subequations}
\label{chiralghost}
\begin{eqnarray}
  \f_\L &\Rightarrow& c \,,\hspace{11ex}
  \f_\L^* \;\Rightarrow\; c \,,
\label{chiralghosta}\\
  \j_\L &\Rightarrow& \sqrt{2} \Sl{A} \h \,,
\label{chiralghostb}\\
  F_\L &\Rightarrow& 2i \bh P_-\l \,,\qquad
  F_\L^* \;\Rightarrow\; 2i \bh P_+\l \,.
\label{chiralghostc}
\end{eqnarray}
\end{subequations}
The virtue of this replacement is that, under the action of the off-shell
BRST operator $s^{off}$,
the fields on the right-hand side of Eq.~(\ref{chiralghost})
transform into each other in the same way as the components
of a gauge-singlet matter supermultiplet do,
according to Eq.~(\ref{matterBRST}).\footnote{
  The reader may recognize Eq.~(\ref{chiralghost})
  as the values assigned to the components of the ghost superfield
  when a linear supersymmetry transformation is accompanied by a supergauge
  transformation to restore the Wess--Zumino gauge.
  The value assigned to $\Im \f_\L(x)$ is zero, and no particular value
  is assigned to $\Re \f_\L(x)$, which is merely renamed as $c(x)$.
}

Having found what shape the ghost superfield takes in
the off-shell formalism, it is now straightforward to obtain the
highest component of the product superfield $\L\F$, assuming $\F=(\f,\j,F)$
is a gauge singlet. Explicitly, it reads
\begin{equation}
  \db^{off} =
  2 \f \bh P_- \l
  +\sqrt{2} \bh \Sl{A} P_+ \j
  -(i/g)Fc \,.
\label{Doff}
\end{equation}
This object transforms into a total derivative under the action of $s^{off}$,
which is expected since it is the highest component of a composite
chiral supermultiplet that behaves as a gauge singlet.
The further transition to the on-shell formalism
is done using Eq.~(\ref{dondef}).\footnote{
  It can be checked that all the steps needed to verify on-shell closedness
  go through (see discussion below Eq.~(\ref{dondef})).
}

Once again, we may take $\F$ in Eq.~(\ref{Doff}) to be a (dimensionful) constant.
For $(\f,\j,F)=(\cc/2,0,0)$, where $\cc$ is given by Eq.~(\ref{photinoanb}),
we reproduce the first term on the right-hand side of Eq.~(\ref{photinoana}).

Equation~(\ref{Doff}) is the key to understanding why spurionization
helps us rule out the abelian gaugino anomaly.  The reason is \textit{not}
that the cohomology becomes trivial; indeed, Eq.~(\ref{Doff}) represents a
nontrivial cohomology class
in the spurionized theory.  What rules out Eq.~(\ref{Doff})
in the spurionized theory is that it depends on $c(x)$
(and not on $\partial_\m c(x))$, and such dependence is not allowed,
as reviewed in Sec.~\ref{mpw1rev}.

\section{\label{ARgaugefix} Supersymmetric local Ward identities}
In this appendix we derive the local version of supersymmetric
Ward identities.  As usual, this is done by promoting the global
supersymmetry parameter $\h$ to a local field.  We assume that,
when $\h$ is still a global parameter, a choice of the
symmetry-restoring counterterms exists such that the
breaking term~(\ref{Dbreaking}) vanishes to all orders in
perturbation theory.

We begin by re-expressing the ST identity\footnote{
  For a related and more detailed discussion,
  see the appendix of Ref.~\cite{MSR}.
}
in terms of the renormalized connected functional $\wr^{(n)}=\wr^{(n)}(J_I,K_I)$,
\begin{equation}
  -\sum_I \int d^4x\, J_I(x) \frac{\d\wr^{(n)}}{\d K_I(x)}
  +\sum_j \frac{\partial\wr^{(n)}}{\partial\z_j}
            \frac{\partial\wr^{(n)}}{\partial k_j} = 0 \,.
\label{STW}
\end{equation}
We have used that each field $\f_I(x)$ is an effective field
associated with the dynamical field $\F_I(x)$, with
$J_I(x) = - \d \G_r^{(n)}/ \d \f_I(x)$ being the corresponding source field.
While this is entirely standard, it is important to notice that Eq.~(\ref{STW})
would \textit{not} be valid in the presence of external spurion fields.
The $J$ sources couple to dynamical fields, and the Legendre transform
back to the connected functional exists for dynamical fields only.

We now promote the supersymmetry parameter to a local field $\h(x)$.
This leads to the appearance of a new term $\propto \pa\m\h(x)$
that we will discuss shortly.
We differentiate the new identity with respect
to $\h(x)$, followed by setting to zero $\h(x)$, as well as the
other BRST parameters $\c$ and $\x_\m$, and the $K$ sources
that couple to the BRST variations.  The result is
\begin{subequations}
\label{localSTconn}
\begin{eqnarray}
  \pa\m S_{\m\a}^{(n)}(x) \cdot \wr^{(n)}
  &=& \sum_I \int d^4y\,
  J_I(y) \frac{\d}{\d\bh_\a(x)} \frac{\d \wr^{(n)}}{\d K_I(y)}
\label{localSTconna}\\
  &=&  -\sum_I \int d^4y\,
        J_I(y) \frac{\d}{\d\bh_\a(x)} \ts_d^{(n)} \F_I(y) \cdot \wr^{(n)} \,.
\label{localSTconnb}
\end{eqnarray}
\end{subequations}
The notation $\co\cdot \wr$ now stands for an insertion of $\co$
into a connected diagram if $\co$ is composite,
whereas if $\co$ is linear in the dynamical fields, it
means that $\co$ is an external leg of the connected diagram.
It is understood that we set $\h(x)=\c=\x_\m=K_I(x)=0$
after taking the functional derivatives.  Using Eq.~(\ref{paron}), it follows that
the term containing the variation of the parameters $\z_i$
in Eq.~(\ref{STW}) drops out.  In arriving at Eq.~(\ref{localSTconnb})
we have made use of the Regularized Action Principle, where
the renormalized transformation $\ts_d^{(n)}$ is still defined by Eq.~(\ref{ts}).
Having performed the functional differentiation with respect to the $K$
sources, we may set $S_{ext,d}^{(n)}=0$ in Eq.~(\ref{localSTconnb}).

The insertion of the renormalized supersymmetry current $S_{\m\a}^{(n)}$
on the left-hand side of Eq.~(\ref{localSTconn}) originates from
an insertion of $\ts_d^{(n)} S_d^{(n)}$ in the regularized theory.
At the classical level, when $\h$ is promoted to a local field
the variation of the action gives rise
to $-\pa\m\h$ times the classical supersymmetry current.
In the hypothetical case that the regularization preserved all the classical
symmetries, $\ts_d^{(n)} S_d^{(n)}$ would still take the form of $-\pa\m\h$
times a (regularized) current.
In reality,  $\ts_d^{(n)} S_d^{(n)}$ includes terms that do not vanish
for constant $\h$, originating from the explicit breaking of (some of) the
classical symmetries by the regularization.
We have assumed that, after removing the cutoff, the breaking term
is cancelled by symmetry-restoring counterterms.
Namely, when $\h$ is a global parameter,
after adding the symmetry-restoring counterterms the breaking term
is the integral of a total derivative.  When the supersymmetry
parameter is promoted to a local field, any such total-derivative terms
become proportional to $\pa\m\h$, and are absorbed into
the renormalized supersymmetry current.

We next perform the functional variation with respect
to $\h(x)$ on the right-hand side of Eq.~(\ref{localSTconn}),
followed by setting $\h(x)=0$.  This gives
\begin{eqnarray}
  \pa\m S_{\m\a}^{(n)}(x) \cdot \wr^{(n)}
  &=&
  \sum_I \bigg( -J_I(x) \bd_\a^{(n)} \F_I(x)
\label{connWI}\\
  && + X_\a^{(n)}(x) \int d^4y\,J_I(y) \ts_g^{(n)} \F_I(y) \bigg)
  \cdot \wr^{(n)} \,,
\nonumber
\end{eqnarray}
where
\begin{equation}
   X_\a^{(n)} = Z_{\cbar\l}^{(n)} (\pa\m \cbar_a) (\g_\m\l_a)_\a \,,
\label{Xalpha}
\end{equation}
and $Z_{\cbar\l}^{(n)}$ is a wave-function renormalization constant.\footnote{
  $\bh X^{(n)}$ is the only source-field independent term in the renormalized
  action that is linear in $\h$.
  To show this, we use, in addition to the ghost number
  and dimensions of the fields, that the renormalized
  action depends on $\cbar$ only through $\pa\m\cbar$,
  which, in turn, follows from a shift symmetry, $\cbar \to \cbar + const$,
  of the classical action (see also Sec.~\ref{mpw1rev}).
}
Now $\wr^{(n)}=\wr^{(n)}(J_I)$,
as all other external fields and BRST parameters
have been set to zero.  The parameter-less renormalized supersymmetry
transformation is defined by
\begin{equation}
  \bd_\a^{(n)} \F_I(x) = \frac{\partial}{\partial \bh_\a}\;
  \ts^{(n)}_d \F_I(x) \,,
\label{dsusy}
\end{equation}
(in this equation we take $\h$ to be global).\footnote{
  The classical transformation $\bd_\a^{(0)}$ vanishes when acting
  on ghost-sector fields, because there are no terms linear in $\h$
  in Eq.~(\ref{gfBRST}).
}

The first term on the right-hand side of Eq.~(\ref{connWI}) is the familiar
contact term
that generates the (renormalized, parameter-less) supersymmetry transformation
of each dynamical field.  The last term is unusual.
It arises because of the explicit dependence of the extended gauge-fixing
action~(\ref{Segf}) on the supersymmetry parameter.  Since the factor of $\h$
is here provided by the extended gauge-fixing action,
we pick up from $\ts_d^{(n)}$ only
the part that survives when setting $\h(x)=\x_\m=\c=0$.
By definition, this is $\ts_g^{(n)}$.

Finally, we perform the functional differentiations with respect to the
$J$ sources and then set them to zero, obtaining the
supersymmetric local Ward identity\footnote{
  Generalizations that involve composite operators
  may be found in the literature.
}
\begin{eqnarray}
  \frac{\partial}{\partial x_\m}
  \svev{S_{\a\m}^{(n)}(x)\, \F_{I_1}(y_1) \cdots \F_{I_k}(y_k) }
  &=&
\label{gfWIgen}\\
  && \hspace{-20ex} = \ \
  \sum_{j=1}^k \d^4(x-y_j)
  \svev{\F_{I_1}(y_1) \cdots \bd_\a^{(n)} \F_{I_j}(y_j) \cdots \F_{I_k}(y_k)}
\NON
  && \hspace{-17ex}
  + \svev{(\ts_g^{(n)}\, X_\a^{(n)}(x)) \F_{I_1}(y_1) \cdots \F_{I_k}(y_k)} \,.
\nonumber
\end{eqnarray}
In going from the last term of Eq.~(\ref{connWI}) to the last term
of Eq.~(\ref{gfWIgen}) we used the invariance of the theory
under the renormalized gauge-BRST transformation $\ts_g^{(n)}$.
The last term in Eq.~(\ref{gfWIgen}) vanishes when applying LSZ reduction,
leading to the familiar result that the perturbative $S$ matrix
is supersymmetric.

We conclude by stressing an obvious but important fact.
All terms in the local Ward identity originate from varying dynamical fields.
There are no terms that arise from varying parameters and/or
external fields.
This is as it should be, since a Ward identity is derived from the invariance
of the path integral under a change of integration variables,
\ie, under a transformation of the dynamical fields only.

\section{\label{MPW2spr} The MPW2 spurions}
In this appendix we show that by introducing the spurion fields
of Ref.~\cite{MPW2} (henceforth denoted MPW2) one cannot rule out
the abelian gaugino anomaly of Sec.~\ref{photino}.
We will use the example of the
superpotential introduced in Eq.~(\ref{Wmass}), which involves a single mass
parameter $m$.  We refer to Sec.~\ref{dyntheory} for notation specific
to that example.

MPW2 introduce a doublet $(u,v)$ of external spurion fields
with transformation rules that, in our notation, read\footnote{
  Notice that if we disregard the $R$ transformation part, this is the same
  structure as found for the ghost-sector doublet $(\cbar,-ib)$.
}
\begin{subequations}
\label{dblt}
\begin{eqnarray}
  s u &=& v + \x_\n\pa{\n} u  + i(2/3)\c u\,,
\label{dblta}\\
  s v &=& \bh\g_\n\h\,\pa{\n} u + \x_\n\pa{\n} v + i(2/3)\c v \,.
\label{dbltb}
\end{eqnarray}
\end{subequations}
We require that the spurionized action reduces to the original action
when the spurion fields take the constant values
$u(x)=0$ and $v(x)=m$.  This correspondence dictates that $v$ is bosonic,
with mass dimension one and ghost-number zero,
whereas $u$ is Grassmann, with mass dimension zero and ghost-number $-1$.
In addition, $u$ and $v$ must both have the same $R$ charge as $m$.

As a warm-up, let us discuss the construction of the spurionized classical
action.  It should be remembered that, unlike in Ref.~\cite{MPW2}, here
the classical action is supersymmetric from the outset; without any spurions,
the on-shell action satisfies the ST identity.
Of course, we require that this remains
true after the introduction of the spurion fields.

In App.~\ref{offshell}, given an off-shell BRST invariant
classical action, we have shown how to construct the corresponding
on-shell action that satisfies the classical ST identity.
It is therefore enough to construct the spurionized off-shell action.
This case is simple, because the $m$ dependence is contained in $S_m$,
where
\begin{subequations}
\label{Sm}
\begin{eqnarray}
  S_m &=&  -i \int d^4x (m \cf + m^* \cf^*)\,,
\label{Sma}\\
  \cf &=& \f_+ F_- + \f_- F_+ + i \bj_+ P_+ \j_- \,,
\label{Smb}\\
  \cf^* &=& \f_+^* F_-^* + \f_-^* F_+^* + i \bj_+ P_- \j_- \,,
\label{Smc}
\end{eqnarray}
\end{subequations}
which is BRST invariant all by itself.
The corresponding spurionized off-shell action is
\begin{subequations}
\label{SMPW2}
\begin{eqnarray}
  S_{(u,v)} &=& -i s^{off} \int d^4x (u \cf + u^* \cf^*)
\label{SMPW2a}\\
  &=& \int d^4x \left( -i v \cf -i v^* \cf^*
  + \sqrt{2} \bh u_5^* \slash\partial (\f_{5+} \j_- + \f_{5-} \j_+) \right) \,.
\label{SMPW2b}
\end{eqnarray}
\end{subequations}
Since the off-shell BRST operator $s^{off}$ is nilpotent,
this action is manifestly BRST invariant.
From Eq.~(\ref{SMPW2b}) it follows that $S_{(u,v)}$ indeed reduces to $S_m$
when the spurions take the constant values prescribed above.

The spurionized on-shell action is obtained as usual via Eq.~(\ref{Sonshell}).
Observe that the dependence of Eq.~(\ref{SMPW2b}) on the auxiliary fields $F_\pm$
is the same as in Eq.~(\ref{Sma}), except for the replacement $m \to v$.
It follows that the spurionized on-shell action is obtained by
substituting $v$ for $m$ everywhere, and adding the $u$-dependent
terms from Eq.~(\ref{SMPW2b}).

\medskip
We now turn to the main point of this appendix.  According to
App.~\ref{ARgaugefix}, the continuity equation
for a conserved supersymmetry current is
\begin{equation}
  \pa\m S_{\m\a}^{(n)} = \ts_g^{(n)} X_\a^{(n)} \,.
\label{contScrnt}
\end{equation}
The right-hand side, which originates from the gauge-fixing procedure,
is gauge-BRST exact, and vanishes on the physical Hilbert space.

Let us now assume the existence of an abelian gaugino anomaly.
For definiteness we assume that, at order $n$ in perturbation theory,
there is a choice of the counterterms that brings the breaking term
into the following form
\begin{equation}
  \D^{(n)} = c \int d^4x\, \bh m_5^* \O_5^* \l \,,
\label{deltam}
\end{equation}
with $c=c^*\ne 0$.  The parameter $m$ was introduced already,
while the dimension-two parameter $\O$ originates from the linear part
of the superpotential.  By repeating the steps of App.~\ref{ARgaugefix}
in the presence of the breaking term~(\ref{deltam}), we find
the anomalous divergence equation
\begin{equation}
  \pa\m S_{\m\a}^{(n)} = \ts_g^{(n)} X_\a^{(n)} + c m_5^* \O_5^* \l  \,.
\label{noconserve}
\end{equation}

Of course, it may be that the assumption we have just made is false.
In fact, in the main text we proved that
the abelian gaugino anomaly~(\ref{photinoan}) will never arise.
However, the question here is whether the same conclusion can be drawn
by introducing the spurion doublet $(u,v)$.

As we will now show, already at the one-loop level the answer
is on the negative. To avoid irrelevant technical issues,
we assume that the spurions have
constant but otherwise arbitrary values $u(x)=u_0$, $v(x)=v_0$.\footnote{
  Clearly, any counterterm that depends on $\pa\m u$ and/or $\pa\m v$
  cannot eliminate the anomalous-divergence term $c v_5^* \O_5^* \l$
  in Eq.~(\ref{spurnocons}) below.
}
We start with a set of one-loop counterterms for the spurionized theory
obtained by substituting $m \to v_0$ in the one-loop counterterms
of the original theory.  At this point, none of the counterterms
depend on $u_0$.  Now, according to MPW2, in the theory with
the external spurion doublet the breaking term must always be $\cb$-exact.
In the case at hand, we must therefore have\footnote{
  We use that, as follows from Eq.~(\ref{dbltb}),
  the transformation rule of the constant mode $v_0$
  is the same as that of $m$ (Eq.~(\ref{parone})).
}
\begin{eqnarray}
  \D^{(n)} &=& c \int d^4x\,
  \left( \bh v_5^* \O_5^* \l - u_5^* \O_5^* \cb \l \right)
\label{deltav}\\
  &=&  c\, \cb \int d^4x\, \bh u_5^* \O_5^* \l  \,,
\nonumber
\end{eqnarray}
with now $u_5 = P_+ u_0 + P_- u^*_0$, \etc,
which reduces to Eq.~(\ref{deltam}) if we set $u_0=0$ and $v_0=m$.
In agreement with the observations of MPW2,
we may now eliminate the remaining breaking term by
adding the further counterterm
\begin{equation}
  Q = -c \int d^4x\, \bh u_5^* \O_5^* \l \,,
\label{ctspur}
\end{equation}
thereby restoring the ST identity.

Nevertheless, the introduction of the counterterm $Q$
does \textit{not} remove the anomalous divergence of the supersymmetry current.
Adding a counterterm $bQ$, with $Q$ given by Eq.~(\ref{ctspur}) and $b$ arbitrary,
leads to the following anomalous divergence equation
in the spurionized theory
\begin{equation}
  \pa\m S_{\m\a}^{(n)} = s_g^{(n)} X_\a^{(n)} + c v_5^* \O_5^* \l
  + (b-1)c u_5^* \O_5^* \cb \l \,.
\label{spurnocons}
\end{equation}
We see that the coefficient of $u_5^* \O_5^* \cb \l$ is affected
by the counterterm, whereas the coefficient of $v_5^* \O_5^* \l$ is not.
As already noted in App.~\ref{ARgaugefix},
the reason is that Ward identities are derived by varying
only the dynamical fields.
The only dynamical field on which $Q$ depends is $\l$, and by varying $\l$
we obtain the $b$-dependent term on the right-hand side.
When we despurionize by setting
$u(x)=u_0=0$, $v(x)=v_0=m$, Eq.~(\ref{spurnocons}) reduces back to
the original anomalous-divergence equation~(\ref{noconserve}).

The conclusion is that, by introducing the spurion doublet $(u,v)$,
one \textit{cannot} rule out the existence of an abelian gaugino anomaly.
In contrast, in Sec. 4 we introduced a different spurion framework
with which such an anomaly can, in fact, be ruled out.


\end{document}